\documentstyle[12pt]{article}
\newskip\humongous \humongous=0pt plus 1000pt minus 1000pt

\newif\ifdtup

\def\abs#1{\left| #1\right|}
\def\pr#1{#1^\prime}

\def\beq{\begin{equation}}
\def\eeq{\end{equation}}

\def\beqn{\begin{eqnarray}}
\def\eeqn{\end{eqnarray}}
\relax

\def\dotx{\dotx{\dot\overline{x}}}

\relax

\catcode`\@=11

\@addtoreset{equation}{section}
\def\theequation{\thesection\arabic{equation}}

\def\@normalsize{\@setsize\normalsize{15pt}\xiipt\@xiipt
\abovedisplayskip 14pt plus3pt minus3pt%
\belowdisplayskip \abovedisplayskip
\abovedisplayshortskip \z@ plus3pt%
\belowdisplayshortskip 7pt plus3.5pt minus0pt}

\def\small{\@setsize\small{13.6pt}\xipt\@xipt
\abovedisplayskip 13pt plus3pt minus3pt%
\belowdisplayskip \abovedisplayskip
\abovedisplayshortskip \z@ plus3pt%
\belowdisplayshortskip 7pt plus3.5pt minus0pt
\def\@listi{\parsep 4.5pt plus 2pt minus 1pt
     \itemsep \parsep
     \topsep 9pt plus 3pt minus 3pt}}

\@twosidetrue

\relax

\catcode`@=12

\catcode`\@=11

\def\section{\@startsection{section}{1}{\z@}{3.5ex plus 1ex minus
   .2ex}{2.3ex plus .2ex}{\large\bf}}

\def\thesection{\arabic{section}.}

\def\appendix{\setcounter{section}{0}
 \def\thesection{APPENDIX \Alph{section}:}
 \def\theequation{\Alph{section}.\arabic{equation}}}

\def\ps@headings{\def\@oddfoot{}\def\@evenfoot{}
\def\@oddhead{\hbox{}\hfill
 \makebox[.5\textwidth]{\raggedright\ignorespaces --\thepage{}--
 \hfill {}}}  
\def\@evenhead{\@oddhead}
\def\subsectionmark##1{\markboth{##1}{}}
}

\ps@headings

\catcode`\@=12

\def\figcap{\section*{Figure Captions\markboth
 {FIGURECAPTIONS}{FIGURECAPTIONS}}\list
 {Fig. \arabic{enumi}:\hfill}{\settowidth\labelwidth{Fig. 999:}
 \leftmargin\labelwidth
 \advance\leftmargin\labelsep\usecounter{enumi}}}
 \relax
\def\tablecap{\section*{Table Captions\markboth
 {TABLECAPTIONS}{TABLECAPTIONS}}\list
 {Table \arabic{enumi}:\hfill}{\settowidth\labelwidth{Table 999:}
 \leftmargin\labelwidth
 \advance\leftmargin\labelsep\usecounter{enumi}}}
 \relax
\def\reflist{\section*{References\markboth
 {REFLIST}{REFLIST}}\list
 {[\arabic{enumi}]\hfill}{\settowidth\labelwidth{[999]}
 \leftmargin\labelwidth
 \advance\leftmargin\labelsep\usecounter{enumi}}}
 \relax

\catcode`\@=11

\def\ps@headings{\def\@oddfoot{}\def\@evenfoot{}
\def\@oddhead{\hbox{}\hfill
 \makebox[.5\textwidth]{\raggedright\ignorespaces --\thepage{}--
 \hfill {}}}    
\def\@evenhead{\@oddhead}
\def\subsectionmark##1{\markboth{##1}{}}
}

\ps@headings

\relax

\relax
\def\pl#1#2#3{{\it Phys. Lett. }{\bf #1}(19#2)#3}

\def\prl#1#2#3{{\it Phys. Rev. Lett. }{\bf #1}(19#2)#3}

\def\pr#1#2#3{{\it Phys. Rev. }{\bf #1}(19#2)#3}
\def\np#1#2#3{{\it Nucl. Phys. }{\bf #1}(19#2)#3}

\relax

\begin{document}
\input epsfig.sty
\newcommand\sss{\scriptscriptstyle}
\newcommand\as{\alpha_{\sss S}}
\newsavebox\tmpfig
\newcommand\settmpfig[1]{\sbox{\tmpfig}{\mbox{\ref{#1}}}}
\newsavebox\tttfig
\newcommand\settttfig[1]{\sbox{\tmpfig}{\mbox{\ref{#1}}}}
  \newcommand{\ccaption}[2]{
    \begin{center}
    \parbox{0.85\textwidth}{
      \caption[#1]{\small{\it{#2}}}
      }
    \end{center}
    }
\begin{titlepage}
\nopagebreak
\vspace*{-1in}
{\leftskip 11cm
\normalsize
\noindent
\newline
hep-ph/9501240\\
CERN-TH.7537/94\\
GeF-TH-1/95\\
IFUM 489/FT\\

}
\vfill
\begin{center}
{\large \bf A Study of Ultraviolet Renormalon Ambiguities}

{\large \bf in the Determination of $\as$ from $\tau$ Decay}
\vfill
{\bf G. Altarelli}, {\bf P. Nason\footnotemark}
\footnotetext{On leave of absence from INFN, Sezione di Milano, Milan, Italy.}
and
{\bf G. Ridolfi\footnotemark}
\footnotetext{On leave of absence from INFN, Sezione di Genova, Genoa, Italy.}
\vskip .3cm
{CERN TH-Division, CH-1211 Geneva 23, Switzerland}
\end{center}
\vfill
\nopagebreak
\begin{abstract}
{\small
The divergent large-order behaviour of the perturbative series relevant
for the determination of $\as$ from $\tau$ decay is controlled
by the leading ultraviolet (UV) renormalon. Even in the absence of the
first infrared (IR) renormalon, an ambiguity of order
$\Lambda^2/m_\tau^2$ is introduced. We make a quantitative study of the
practical implications of this ambiguity. We discuss the magnitude of UV
renormalon corrections obtained in the large-$N_f$ limit, which, although
unrealistic, is nevertheless interesting to some extent. We then study a number
of improved approximants for the perturbative series, based on a change of
variable in the Borel representation, such as to displace the leading
UV renormalon singularity at a larger distance from the origin than the first
IR renormalon. The spread of the resulting values of $\as(m^2_\tau)$ obtained
by different approximants, at different renormalization scales, is exhibited
as a measure of the underlying ambiguities. Finally, on the basis of
mathematical models, we discuss the prospects of an actual improvement, given
the signs and magnitudes of the computed coefficients, the size
of $\as(m^2_\tau)$ and what is known of the asymptotic properties of the
series.
Our conclusion is that a realistic estimate of the theoretical error cannot go
below $\delta\as(m^2_\tau) \sim \pm 0.060$, or
$\delta\as(m^2_{\sss Z}) \sim \pm 0.006$.
}
\end{abstract}
\vfill
CERN-TH.7537/94
\newline
December 1994    \hfill
\end{titlepage}

\section{Introduction}
The possibility of measuring $\as$ from $\tau$-decay has been extensively
studied in a series of interesting papers, in particular by Braaten,
Narison and Pich [\ref{SchilcherTran}-\ref{DiberderPich}].
The relevant quantity is
$R_\tau= \Gamma(\tau \to \nu_\tau +
{\rm hadrons})/\Gamma(\tau \to \nu_\tau +l\nu)$,
with $l=e,\mu$. At present [\ref{Duflot}] the ALEPH
collaboration finds
\beq
{(R_\tau)}_{exp} = 3.645 \pm 0.024.
\eeq
Based on this result, it is argued,
if the formalism of QCD sum rules is assumed, according to
SVZ [\ref{SVZ}], that
\beq
\as(m^2_\tau)=0.355 \pm 0.021
\eeq
and finally
\beq
\as(m^2_{\sss Z})=0.121 \pm 0.0016 (exp) \pm 0.0018 (th)=0.121 \pm 0.0024.
\eeq
Given that $m_\tau$ is so small, this
determination of $\as(m^2_{\sss Z})$ appears [\ref{Pumplin}-\ref{Truong}]
a bit too precise!

In defence of this method [\ref{Narison1}] one can certainly point out that
$R_\tau$ has several combined advantages. Dropping some inessential
complications, $R_\tau$ is an integral in $s$ of a spectral function $R(s)$
which is the analogue of $R_{e^+e^-}(s)$ but for the case of charged weak
currents. Thus, first, $R_\tau$ is even more inclusive than $R_{e^+e^-}(s)$ and
one expects that the asymptotic regime is more precocious for more
inclusive quantities. Second, one can use analyticity in order to
transform the relevant integral into an integral over the circle
$|s| = m_\tau^2$ [\ref{DiberderPich}]. This not only gives some confidence
that the appropriate scale of energy for the evaluation of $R_\tau$
is of order $m_\tau$, but also shows that the integration over the
low-energy domain helps very much in smearing out the complicated behaviour
in the resonance region. Also important is the presence of a phase-space
factor that kills the sensitivity of the spectral function near
${\rm Re}\,s=m_\tau^2$, where there is a gap of validity of the asymptotic
approximations due to the vicinity of the cut singularities and also to the
nearby charm threshold. On the circle $|s|=m_\tau^2$,
asymptotic formulae should be approximately valid for the correlator.
The perturbative component of
$R(s)$ is known up to terms of order $\as(m^2_\tau)^3$
[\ref{GKL}]. One can hope to get some control of the non-perturbative
corrections by using the operator product expansion and some estimate
(either experimental or by some model) of the dominant condensates,
in the spirit of the QCD sum rules [\ref{SVZ}].

This series of virtues of $R_\tau$ is indeed real but would not be
sufficient in itself to justify the precision on $\as(m^2_\tau)$ which is
claimed. The real point is that no corrections of order $1/m_\tau^2$
are assumed to exist. The fact that there is no
operator with the corresponding dimension in the short distance
expansion is not sufficient, because there could be non-leading
corrections in the coefficient function of the leading operator. We
think it is a fair statement that there is no theorem that guarantees
the absence of $\Lambda^2/m_\tau^2$ terms in $R_\tau$ in the massless limit;
no theorem that proves that terms of order $\Lambda^2/m_\tau^2$
cannot arise from the mechanism that generates confinement. But even if
{\it in principle} the above theorem would exist,
still, {\it in practice}, there would be ambiguities on the leading-term
perturbative expansion of order $\Lambda^2/m_\tau^2$ from the ultraviolet
renormalon sequence associated to the divergence of the perturbative
series for the spectral function [\ref{tHooft}-\ref{Lovett}].
The present note is mainly devoted to a quantitative discussion
of the impact of UV renormalon ambiguities on the determination of
$\as(m^2_\tau)$. On the basis of the accumulated knowledge on renormalon
behaviour, we address the question of what is the theoretical error
on $\as(m^2_\tau)$ and examine possible ways to decrease it. We discuss the
magnitude of UV renormalon corrections obtained from explicit
calculations [\ref{Zakharov}-\ref{Lovett}],
which although based on unrealistic simplified
schemes, are nevertheless interesting to some extent.  We then study
a number of improved approximants for the perturbative series, based
on a change of variable in the Borel representation
[\ref{Lautrup},\ref{Mueller1}], such as
to displace the leading UV renormalon singularity at a larger
distance from the origin than the first infrared (IR) renormalon. The
spread of the resulting values of $\as(m^2_\tau)$ obtained by different
approximants, at different renormalisation scales is exhibited as a
measure of the underlying ambiguities.
Finally, on the basis of
mathematical models, we discuss the prospects of an actual improvement, given
the signs and the magnitudes of the computed coefficients, the size
of $\as(m^2_\tau)$ and what is known of the asymptotic properties of the
series.
Our conclusion is that a realistic extimate of the theoretical error cannot go
below $\delta\as(m^2_\tau) \sim \pm 0.060$, or $\delta\as(m^2_{\sss Z}) \sim
\pm 0.006$.

The organisation of this article is as follows. In sect.2 we summarise
the basic formulae and discuss different procedures to do the
integration over the circle that differ by resumming or not an
infinite series of ``large $\pi^2$ terms''. We discuss the relative merits
of the various procedures and their scale dependence. In sect.3 we
introduce the problems related to the divergence of the perturbative
series, we review the Borel transform method and the renormalon
singularities. In sect.4 we derive some useful formulae obtained in
the Borel space after integration on the circle. In sect.5 we consider
the explicit form for the leading UV renormalon singularity derived in
perturbation theory in the large $N_f$ limit, $N_f$ being the number of
flavours. This limit is not meant to be realistic, but, for
orientation, we evaluate the quantitative impact that such an UV
renormalon would have on the determination of $\as(m^2_\tau)$. We find that
this effect is rather small. In sects.6,7, which contain the main
original results of this work, we introduce and study a number of
improved approximants that could in principle suppress the ambiguity
from the leading UV renormalon. We study the combined effects of
different, a priori equivalent, procedures,  different accelerators of
convergence and different choices of the renormalisation scale. We
also study in a simple mathematical model under which conditions for
the known coefficients of the series the accelerator method leads to a
better approximation of the true result. Finally, in sect. 8 we
present our conclusion.

\section{Basic Formulae and Truncation Ambiguities}
The quantity of interest is the integral over the hadronic squared
mass $s$ in $\tau$ decay of a function $R(s)$ analogous to $R_{e^+e^-}(s)$,
weighted by a phase-space factor. In the limit of massless $u,d,s$
quarks we have [\ref{SchilcherTran}-\ref{LuoMarciano}]:
\beq
\label{rtau1}
R_\tau=\int_0^{m_\tau^2} \frac{ds}{m_\tau^2}
2\left(1-\frac{s}{m_\tau^2}\right)^2
\left(1+\frac{2s}{m_\tau^2}\right)R(s).
\eeq
$R(s)$ is proportional to the imaginary part of a current-current correlator:
\beq
R(s)=\frac{N}{\pi}{\rm Im}\, \Pi(s)=\frac{N}{2\pi i}
\left[\Pi(s+i\epsilon)-\Pi(s-i\epsilon)\right].
\eeq
The normalization factor $N$ is defined in such a way that, in zeroth
order in perturbation theory, $R(s)=3$.
In turn, the correlator $\Pi(s)$ is related to the Adler function
$D_\tau(s)$, defined in such a way as to remove a constant:
\beq
D_\tau(s)=-s\frac{d}{ds} N\Pi(s).
\eeq
By first integrating by parts and then using the Cauchy theorem one
obtains for $R_\tau$ the result
\beq
\label{rtau2}
R_\tau=\frac{1}{2\pi i}
\oint_{|s|=m_\tau^2} \frac{ds}{s}
\left(1-\frac{s}{m_\tau^2}\right)^3
\left(1+\frac{s}{m_\tau^2}\right)D_\tau(s).
\eeq
The Adler function $D_\tau$ has a perturbative expansion of the form:
\beq
\label{dtauexp}
D_\tau(s)=
D^0_\tau \sum_{n=0}^{\infty} D_n a(-s)^n
\simeq
D^0_\tau \left[1+D_1 a(-s)+D_2 a^2(-s)+D_3 a^3(-s)+\ldots\right],
\eeq
where $a=\as/\pi$, $D^0_\tau=3(1+\delta)$ where $\delta$ is a known small
electroweak correction, and, for $N_f=3$ in the $\overline{\rm MS}$ scheme,
\beqn
&&D_1=1 \nonumber\\
&&D_2=\left[\frac{11}{2}-4\zeta(3)\right]\beta+\frac{C_A}{12}
-\frac{C_F}{8}=1.640
\nonumber\\
&&D_3=
\left[\frac{151}{18}-\frac{19}{3}\zeta(3)\right]4\beta^2
+2C_A\left[\frac{31}{6}-\frac{5}{3}\left(\zeta(3)+\zeta(5)\right)\right]\beta
\nonumber\\
&&\phantom{D_3=}
+2C_F\left[\frac{29}{32}-\frac{19}{2}\zeta(3)+10\zeta(5)\right]\beta
+C_A^2\left[-\frac{799}{288}-\zeta(3)\right]
\nonumber\\
&&\phantom{D_3=}
+C_A C_F \left[-\frac{827}{192}+\frac{11}{2}\zeta(3)\right]
+C_F^2\left(-\frac{23}{32}\right)=6.371,
\label{Dcoeff}
\eeqn
where $\zeta(3)=1.20206$ and $\zeta(5)=1.03693$, $C_A=N_C=3$ and
$C_F=(N_C^2-1)/(2N_C)=4/3$.
The quantity $\beta = (11C_A-2N_f)/12$ is the
first beta function coefficient [\ref{GA3}]:
\beq
\mu^2\frac{d a(\mu^2)}{d\mu^2}=\beta\left(a(\mu^2)\right);
\;\;\;\;
\beta\left(a\right)=-\beta a^2(1+\beta^\prime a+\ldots)
\eeq
($\beta=9/4$, $\beta^\prime=16/9$ for $N_f=3$).

The expansion in eq.~(\ref{dtauexp}) defines the Adler function at all
complex $s$ with a cut for $s>0$. In the spacelike region, where $s<0$,
$a(-s)$ is real and given asymptotically by ($\mu^2>0$):
\beq
\frac{1}{a(-s)}=\frac{1}{a(\mu^2)}
+\beta\log\frac{-s}{\mu^2}=\beta\log\frac{-s}{\Lambda^2}.
\eeq
If we want $a(-s)$ at some
complex value of the argument, e.g. $s=-|s|\exp(i\theta)$, we can
use the formula
\beq
\label{aq2}
a(-s)=\frac{a(|s|)}{1+\beta a(|s|)i\theta}
\eeq
where the angle $\theta$ is $-\pi$ on the upper tip of the cut for $s$
real and positive, $+\pi$ on the lower tip and zero on the negative
real axis.
The more accurate two-loop expression is given by
\beq
\label{aq2twoloop}
a(-s)=\frac{a(|s|)}{1+\beta a(|s|)i\theta
+\beta^\prime a(|s|)\log(1+\beta a(|s|)i\theta)}.
\eeq

The expansion for $R(s)$ (for $s$ real and positive)
can be obtained from that of $D_\tau$ by the
relation
\beq
R(s)=\frac{1}{2\pi i} \oint_{|s^\prime|=s}\frac{d s^\prime}{s^\prime}
D_\tau(s^\prime).
\eeq
Performing the integration by using the expansion for $D_\tau$ in
eq.~(\ref{dtauexp}) and the expression in eq.~(\ref{aq2twoloop}) for a complex
argument, one obtains
\beq
R(s)=D^0_\tau\left(1+F_1 a(s)+F_2 a^2(s)+F_3 a^3(s)+\ldots \right),
\eeq
and
\beq
\label{FD}
F_1=D_1; \;\;\; F_2=D_2; \;\;\; F_3=D_3-\frac{\beta^2 \pi^2}{3}.
\eeq
The origin of the $\beta^2 \pi^2/3$ term is easily understood. By using
the one-loop expansion for $a(s)$, eq.~(\ref{aq2}), one gets
\beq
\label{Rs}
R(s)=D^0_\tau\left[1+\frac{1}{2\pi\beta i}\log
\frac{1+i\pi\beta a(s)}{1-i\pi\beta a(s)}+\ldots \right]=
D^0_\tau\left[1+a(s)-\frac{\beta^2\pi^2 a^3(s)}{3}+\ldots\right]
\eeq
We have the following observations on this result.
First, for $N_f=3$, the coefficients $F_{1,2,3}$ in the expansion of
$R(s)$ for $\tau$ decay coincide with those of $R_{e^+e^-}$ because the
potentially different terms proportional to $(\sum Q_i)^2$ , with $Q_i$ being
the quark charges, vanish in this case. Second, we observe that
eq.~(\ref{FD}) is obtained by a truncation
of higher-order terms in the quantity  $\beta^2 a^2 \pi^2 \simeq 0.7$
with $a =a(m^2_\tau)$. Note that a similar problem of truncation arises when
the integration over the circle in eq.~(\ref{rtau2}) is performed. In the
early treatments of this problem (e.g. in ref.~[\ref{Braatenetal}]) the
expression of $a(-s)$, which appears on the circle, is taken from
eq.~(\ref{aq2twoloop}) and expanded in $a$ consistently to the order
$a^3$. With this procedure, one obtains
\beq
\label{rtau3}
R_\tau^{({\rm BNP})}=D^0_\tau\left(1+H_1\, a(m^2_\tau)+H_2\,a^2(m^2_\tau)
+H_3\,a^3(m^2_\tau)+\ldots\right),
\eeq
with
\beq
H_1=1;  \quad H_2=5.2023; \quad H_3=26.3666.
\eeq
We will refer to this result as BNP formula (for the authors of
ref.~[\ref{Braatenetal}]). More recently in
ref.~[\ref{DiberderPich}] it was advocated that a
better procedure for performing the integration on the circle is to
keep the full three-loop expression for $a(-|s|e^{i\theta})$, according to
the formula
\beq
\label{rtaupich}
R_\tau^{({\rm LP})}=\frac{D^0_\tau}{2\pi i}
\oint_{|z|=1} \frac{dz}{z}(1-z)^3(1+z)
\left[1+D_1 a(-zm^2_\tau)+D_2 a^2(-zm^2_\tau)
+D_3 a^3(-zm^2_\tau)+\ldots\right].
\eeq
It is this procedure which is currently adopted (LeDiberder-Pich, or LP
method). At fixed experimental value of $R_\tau$, the two procedures lead to
values of $\as(m^2_\tau)$ that differ by terms of order
$\delta\as(m^2_\tau) \sim (\beta^2 a^2 \pi^2) a^2 \sim 0.01$
or $(\beta^2 a^2 \pi^2)^2 a\sim~0.05$. These ``large $\pi^2$ terms''
always arise when one goes from the spacelike to the timelike region
(e.g. similar terms arise [\ref{GA3}] when one relates Drell-Yan processes to
electroproduction or fragmentation functions to structure functions). There
have been many discussions in the past on the opportunity of resumming
these terms [\ref{Pennington}]. If there was a good argument to consider the
expansion for $D_\tau(s)$ in some respect superior to that for $R(s)$ it
could be worthwhile at low energies to keep the expression in
eq.~(\ref{Rs}) in its resummed form rather than to expand in
$\beta^2 a^2 \pi^2$. A glance at eqs.~(\ref{Dcoeff})
shows that, for $n\leq 3$, there are no explicit ``large $\pi^2$ terms''  in
the coefficients $D_n$  of the expansion for $D_\tau$. Similarly when the
integration on the circle is performed with the complete formula for
$a(-|s| e^{i\theta})$ all terms are kept up to order
$a^3(\beta^2 a^2 \pi^2)^n$, i.e. up to
order $a^3$ one expands in $a$ but keeps $\beta^2 a^2 \pi^2$ unexpanded.
Sure enough this sequel of terms exists in reality, so why not take them into
account?  However, the counter-argument is that there are in
perturbation theory terms involving $\pi^2$ that arise from origins other
than the spacelike-timelike connection and, in any case, there are
many terms of the same general magnitude (for example, the term
proportional to $\beta^2$ in $D_3$, eq.~(\ref{Dcoeff})), so that the
advantage of keeping this particular class of terms is likely to be completely
illusory.

One can further consider the scale dependence of the different
procedures. One can write
\beq
\label{rtauBNPscale}
R_\tau^{({\rm BNP})}=D^0_\tau\left(1+\tilde{H}_1\, a(\mu^2)
+\tilde{H}_2\, a^2(\mu^2)+\tilde{H}_3\, a^3(\mu^2)
+\ldots\right),
\eeq
where
\beqn
&&\tilde{H}_1(\mu^2)=H_1
\nonumber \\
&&\tilde{H}_2(\mu^2) = H_2-H_1\beta\log\frac{m_\tau^2}{\mu^2}
\nonumber \\
&&\tilde{H}_3(\mu^2)=H_3-2H_2\beta\log\frac{m_\tau^2}{\mu^2}
+H_1\left[\beta^2\log^2\frac{m_\tau^2}{\mu^2}
-\beta\beta^\prime \log\frac{m_\tau^2}{\mu^2}\right].
\label{Hrescaled}
\eeqn
Analogously, we can study the scale dependence in the LP method. In this case
we have
\beq
\label{rtaupichmu}
R_\tau^{({\rm LP})}=\frac{D^0_\tau}{2\pi i}
\oint_{|z|=1} \frac{dz}{z}(1-z)^3(1+z)
\left[1+\tilde{D}_1 a(-z\mu^2)+\tilde{D}_2 a^2(-z\mu^2)+
\tilde{D}_3 a^3(-z\mu^2)+\ldots\right],
\eeq
where the $\tilde{D}_n$ coefficients are related to the $D_n$ as the
$\tilde{H}_n$ to the $H_n$ in eqs.~(\ref{Hrescaled}).

\begin{figure}[tbhp]
  \begin{center}
    \mbox{
      \epsfig{file=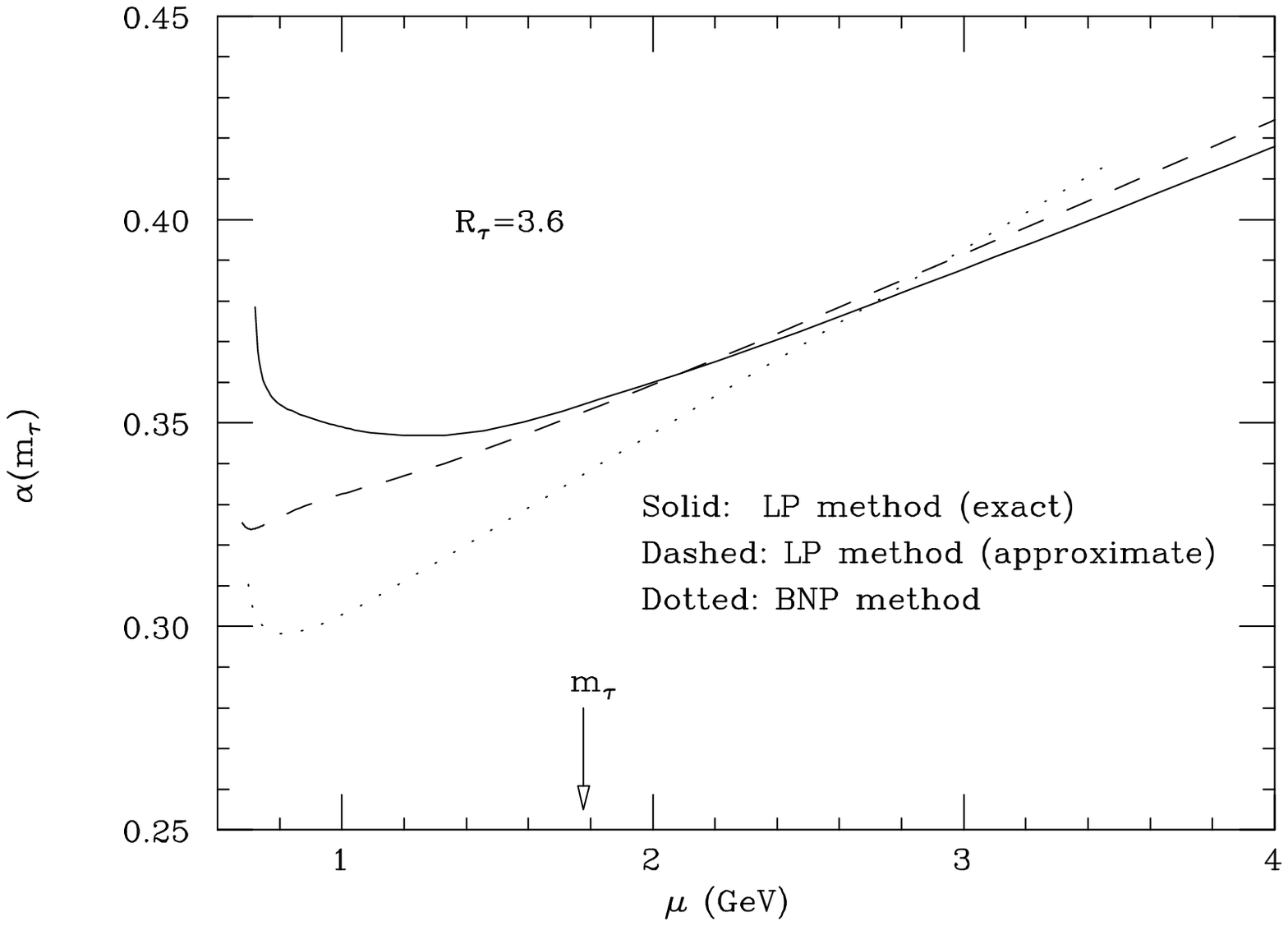,width=0.80\textwidth}
      }
  \ccaption{}{\label{asdet}
The value of $\as(m^2_\tau)$ obtained from $R_\tau=3.6$, as a function
of the renormalization scale $\mu$.}
  \end{center}
\end{figure}

The results of the LP and BNP methods are shown in fig.~\ref{asdet}, where,
assuming a measured value of 3.6 for $R_\tau$, we show the corresponding
determination of $\as(m^2_\tau)$ as a function of the renormalization
scale $\mu$. For comparison, we also show (dashed curve) the determination
obtained with a simplified LP method, in which we use the one-loop
expression eq.~(\ref{aq2}) for $a$.
For $\mu>1$~ GeV there is less $\mu$ dependence if the resummed
expressions are used. This stability is often taken as a possible
indication that resumming is better.

In conclusion, we agree that the resummed formulae provide a less
ambiguous result with respect to a change of scale than the
unresummed expression. However, it is true that a priori it is
not possible to guarantee that a more accurate result is obtained
in one way or the other. As a consequence the spread shown in
fig.~\ref{asdet} for different choices of $\mu$ and of procedure is to
be taken as a real ambiguity. In particular the large discrepancy
at $\mu \sim 1$~GeV is a genuine signal of trouble, especially in view of the
fact that several proposed scale-fixing procedures lead to small values
of $\mu$, e.g. minimal sensitivity, BLM scheme etc. [\ref{maghi}].
A small value of $\mu$ is also suggested by physical considerations,
because the average hadronic mass is well below $m_\tau$. So, on the one hand,
one cannot
sensibly reject the option of small values of $\mu$. On the other hand,
the corresponding value of $a(m^2_\tau)$ becomes very ambiguous at small $\mu$.

\section{Renormalons and Borel Transformation}
\begin{figure}[tbhp]
  \begin{center}
    \mbox{
      \epsfig{file=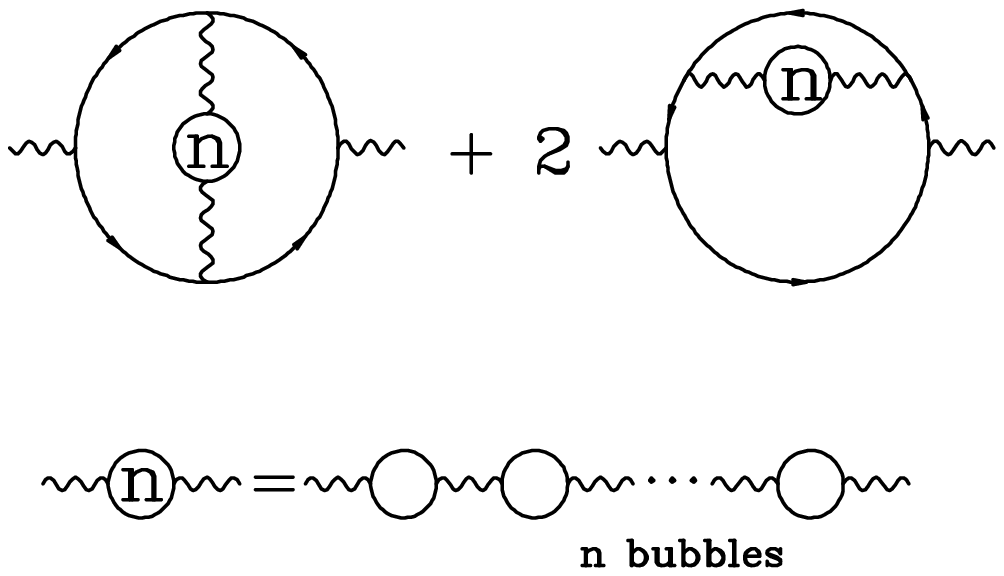,width=0.60\textwidth}
      }
  \ccaption{}{\label{diagrams}
Renormalon diagrams for a two-point correlator.}
  \end{center}
\end{figure}

In the following discussion we first study the properties of $D_\tau(s)$ in
itself, and only later we consider the integration over the circle.
The problem that we now consider is the well known fact that the
series for $D_\tau$ is divergent. Indeed one can identify sequences of
diagrams, depicted in fig.~\ref{diagrams}, called ``renormalon'' terms
[\ref{tHooft}-\ref{Lovett}],
that provide the leading behaviour at large $n$ for the $n$-th coefficient of
the expansion for $D_\tau$.
The renormalon contribution is of the form:
\beq
D_n \sim C_k n! n^{\gamma_k} \left(\frac{\beta}{k}\right)^n
\left[1+{\cal O}(1/n)\right] \;\;\;(n\;{\rm large}).
\eeq
Note the $n!$ behaviour which implies that the series is
divergent. Here, $C_k$ and $\gamma_k\geq 0$ are not known for the real
theory but only, to some extent, in the large-$N_f$ expansion
[\ref{Zakharov}-\ref{Lovett}], and the index $k$ runs over a discrete set
of values:
\beqn
&&k = -1, -2, -3, \ldots \;\;\; {\rm Ultraviolet \;\;(UV)\; \;Renormalons}
\nonumber \\
&&k = +1(?), +2,+3, \ldots \;\;\; {\rm Infrared\;\; (IR)\;\; Renormalons}
\eeqn
In the above list we omit the contribution from instantons,
which appear at rather large values of $k$.
They have been computed in ref.~[\ref{Instantons}] and shown to be
small. The UV or IR renormalons arise from the limits of
large or small virtuality, respectively, for the exchanged
gluon(s). As hinted by the question mark, the $k=+1$ IR renormalon is
probably absent in perturbation theory, but the issue is not really
settled [\ref{West}-\ref{Lovett}]. The absence of this term
is necessary for the consistency of the assumption that all non-perturbative
effects can be absorbed in the condensates. In the following we will
assume that the $k=1$ IR renormalon is indeed absent.

In view of the divergence of the perturbative expansion, one can
possibly give a meaning to the quantity
\beq\label{ddef}
 d(a) = \frac{D_\tau}{D^0_\tau}-1=D_1 a+D_2 a^2+D_3 a^3+\ldots
\eeq
by the Borel transform method [\ref{Hardy}]. One defines the perturbative
expansion of the Borel transform $B(b)$ of $d(a)$ by removing the n! factors:
\beq
\label{boreltr}
B(b)=\sum_{n=0}^\infty D_{n+1}\frac{n^n}{n!} =D_1+D_2 b +D_3\frac{b^2}{2}
+\ldots
\eeq
Then, formally
\beq
\label{antiborel}
d(a)=\int_0^\infty db \,e^{-b/a}B(b)
\eeq
in the sense that the expansion for $B(b)$ reproduces the expansion for
$d(a)$ term by term. What is needed for $d(a)$ to be well defined is that the
integral converges (this cannot be true at all $s$
[\ref{tHooft},\ref{Mueller1}] because of
the singularities of $d(a)$ in the $s$ plane, but this problem can be
neglected in our context) and that $B(b)$ has no singularities in the
integration range. But, as already mentioned, the large-$n$ expansion of
$B(b)$ leads to singularities on the real axis. In fact, at large $n$,
$B(b)$ is essentially given by a geometric series:
\beq
B(b)\sim C_k\sum_n n^{\gamma_k}\left(\frac{\beta b}{k}\right)^n
\sim C_k \Gamma(\gamma_k+1)\left(1-\frac{\beta b}{k}\right)^{-\gamma_k-1}
+ {\rm less \;\; singular \;\; terms}
\eeq
so that it is singular at $b = k/\beta$. Thus the UV renormalons correspond
to singularities at $b = -1/\beta, -2/\beta, \ldots$ and the IR renormalons at
$+2/\beta, +3/\beta,\ldots$. As
a consequence, the convergence radius
of the expansion for $B(b)$ near the $b$-origin is determined by the UV
renormalon at $b = -1/\beta$, independent of the existence of the IR
renormalon at $b = +1/\beta$.

Thus, the perturbative expansion for $B(b)$ can be directly used only to
perform the integration up to $b = +1/\beta$. The contribution from
$b = +1/\beta$ up to $b = \infty$, where the expansion is not valid,
could typically lead to terms of order $\Lambda^2/s$ (or even worse).
For example, if $B(b)$ is sufficiently well behaved at $b = +1/\beta$ and at
$b = \infty$,
\beq
\Delta d(a)=\int_{1/\beta}^\infty db \,e^{-b/a}B(b)
\sim a B(1/\beta) \exp(-1/\beta a)\sim a B(1/\beta) \Lambda^2/s,
\eeq
where we used $a^{-1} \simeq \beta\log(s/\Lambda^2)$ and the fact that the
exponential cuts away all large-$b$ contributions so that $B(b)$ was
approximated by its value near  $b = +1/\beta$. From a different point of view,
at large $n$, the series for $d(a)$ is dominated by the UV renormalon
behaviour with $D_n \sim n^\gamma n! (-\beta)^n$. At fixed small $a$,
the individual terms $|D_n| a^n$ first decrease with $n$, then flatten out
and eventually increase because of the $n!$ factor. The best estimate of the
sum is obtained by stopping at the minimum, for $n\sim n_{opt}$,  given by
$|D_n| a^n \sim |D_{n-1}| a^{n-1}$, or $n_{opt}\sim 1/\beta a$.
{}From the theory of asymptotic series [\ref{Hardy}], the corresponding
uncertainty $\Delta d(a)$ is of order $|D_{n_{opt}}| a^{n_{opt}}$:
\beqn
|D_{n_{opt}}|a^{n_{opt}}
&\sim& \left(1/\beta a\right)^\gamma
\left(1/\beta a\right)! \left(\beta a\right)^{(1/\beta a)}
\nonumber \\
&\sim&
\left(1/\beta a\right)^\gamma
\left(1/\beta a\right)^{(1/\beta a)}e^{(1/\beta a)}
\sqrt{2\pi/\beta a}(\beta a)^{(1/\beta a)}
\nonumber \\
&\sim&
\left(1/\beta a\right)^\gamma
e^{(1/\beta a)} \sqrt{2\pi/\beta a}
\sim \Lambda^2/s \times {\rm logarithms},
\eeqn
(where the Stirling approximation was used: $n!\simeq n^n e^{-n}\sqrt{2\pi
n}$).
Thus one could improve the accuracy of the perturbative expansion by
computing more subleading terms until $n\sim n_{opt}$ is reached and then add a
residual term of order $\Lambda^2/s$. Note that the estimate
$n_{opt}\sim 1/\beta a \sim 4$ indicates a rather small value. However this
estimate is obtained from the behaviour of the leading UV renormalon series,
while there is no alternation of signs and in general no evidence of renormalon
behaviour in the few known terms of the series.

While an accuracy of order $\Lambda^2/s$ is what one gets in practice from the
three-loop expression of $d(a)$, it is true that, in principle, if there
is no IR renormalon at $b=+1/\beta$, $d(a)$ can be better defined. In fact, as
the location of the leading UV renormalon at $b=-1/\beta$ is not in the
integration range, there is the possibility of defining $B(b)$ by
analytic continuation up to $b=+2/\beta$. If this is realised then the
remaining ambiguity, of order $(\Lambda^2/s)^2$, is unavoidable because the
corresponding singularity at $b=+2/\beta$ is on the real axis, so that an
arbitrary procedure to go around it must be defined and the difference
between two such procedures would be of that order. However, since
operators of dimension 4 do exist in the operator expansion, this
ambiguity can be reabsorbed in the non-perturbative condensate terms
[\ref{tHooft}-\ref{David}]. We also understand that the absence of the IR
renormalon at $b=+1/\beta$ is necessary for the consistency of the SVZ
[\ref{SVZ}] approach because  operators of dimension 2 are absent in this
channel. If there would be a singularity at $b=+1/\beta$ the corresponding
ambiguity could not be absorbed in a condensate. There are indeed indications
in
perturbation theory that the first IR renormalon does not appear
[\ref{Beneke},\ref{Lovett}]. However there could be non-perturbative sources
of breaking of the operator expansion at non-leading level. After all no
theory of confinement could be built up from perturbation theory and
renormalons. But, in practice, independent of the existence of the IR
renormalon at $b=+1/\beta$, the accuracy to be expected from the first three
terms in the expansion for $d(a)$, as they have been used so far in the actual
determination of $\as$, is of order $\Lambda^2/s$.

\section{The Integration over the Circle in the Borel Transform Formalism}
\begin{figure}[tbhp]
  \begin{center}
    \mbox{
      \epsfig{file=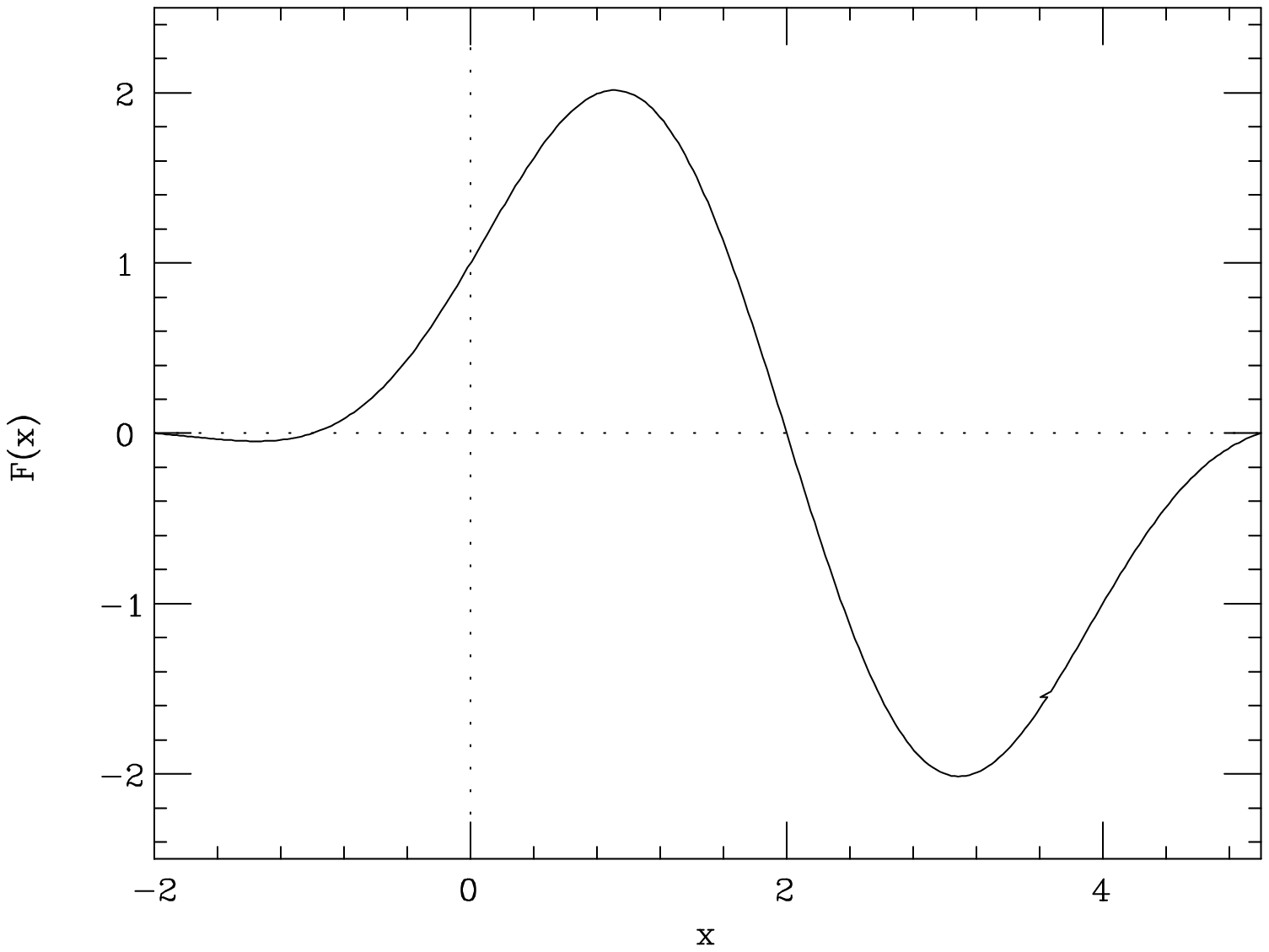,width=0.80\textwidth}
      }
  \ccaption{}{\label{FX}
Plot of the function $F(x)$.
}
  \end{center}
\end{figure}

In this section we show that the integration over the circle in
eq.~(\ref{rtau2}) for $R_\tau$ is particularly simple in the Borel
representation. Starting from eqs.~(\ref{aq2}) and (\ref{antiborel}) we have
\beq
r=\frac{R_\tau}{D_\tau^0}-1=\frac{1}{2\pi i}
\oint_{|s|=m_\tau^2} \frac{ds}{s}
\left(1-\frac{s}{m_\tau^2}\right)^3
\left(1+\frac{s}{m_\tau^2}\right)
\int_0^\infty db \,e^{-b/a(-s)}B(b).
\eeq
We work in the approximation where the two-loop coefficient $\beta^\prime$
in the beta function is neglected. Then, according to eq.~(\ref{aq2}),
we can replace $1/a(-s)$ by $1/a + i\beta\theta$, where $a$
is $a(|s|) = a(m^2_\tau)$, invert the integration order and write
$s/m_\tau^2=-\exp(i\theta)$:
\beq
r=\int_0^\infty db \,e^{-b/a}B(b)\frac{1}{2\pi i}
\int_{-\pi}^\pi id\theta \,(1+e^{i\theta})^3(1-e^{i\theta})
e^{-ib\beta\theta}.
\eeq
The integration is easily performed, with the result
\beqn
r
&=&\int_0^\infty db \,e^{-b/a}B(b)
\frac{-12\sin(\beta b \pi)}{\beta b(\beta b-1)(\beta b-3)(\beta b-4)\pi}
\nonumber\\
&=&\int_0^\infty db \,e^{-b/a}B(b)F(\beta b).
\label{rapp}
\eeqn
We see that, in first approximation, the effect of going from $a(-s)$ to
$a(|s|)$ by integrating over the circle is to multiply the Borel
transform $B(b)$ by the factor $F(\beta b)$. For real $x$, the function $F(x)$
is shown in fig.~\ref{FX}.
It is an entire function in the whole complex plane with a good
behaviour at infinity on the real axis. The factor $F(x)$ has simple
zeros at the location of all UV renormalons and also of all IR
renormalons with the exception of those at $\beta b= 1$ (if any), 3 and
4. Since, in general, the corresponding singularities are not simple
poles, they are not eliminated, but their strength is attenuated
(this point will be discussed in more detail in sect.~5.).
Equation~(\ref{rapp})
obviously coincides with the LP approach [\ref{DiberderPich}] in the limit
$\beta^\prime =\beta^{\prime\prime}=0$. One can also repeat the
procedure by expanding in $a(\mu^2)$ instead of $a(m^2_\tau)$, according
to eq.~(\ref{rtaupichmu}). The resulting
$\mu$ dependence is shown in fig.~\ref{asdet} together with the analogous
results for the BNP [\ref{Braatenetal}] and the LP formulae.

\section{Large-$N_f$ Evaluation of Renormalons and their Resummation}
As well known, the typical renormalon diagrams of QED and QCD can be
evaluated in the large-$N_f$  limit and their structure is simple in this
limit [\ref{Zakharov}-\ref{Lovett}]. In the abelian case the large-$N_f$
limit corresponds to the large-$\beta$ limit, and this is believed to be
true also in the non-abelian gauge theory in spite of the fact that the
beta function in this case cannot be evaluated only in terms of vacuum
polarisation diagrams. The sequence of dominant terms in $\beta^n$ generalises
the terms in $\beta$ and $\beta^2$ that appear in $\tilde{D}_2$ and
$\tilde{D}_3$ respectively. As argued in a recent paper, ref.~[\ref{VZ}], the
determination of the exact behaviour of the UV renormalon series
may be very different from the one indicated in the large-$\beta$ limit.
While we do not know the complete form of the leading UV
renormalon at $b=-1/\beta$, we can nevertheless compute, for orientation,
the quantitative impact of its approximate form at large $\beta$ on the
determination of $a(m^2_\tau)$. From eq.~(42) of ref.~[\ref{Lovett}]  one
obtains the large-$\beta$ expression of the contribution of the leading
UV renormalon at $b=-1/\beta$ to the Borel transform $B(b)$:
\beq
B(b)=\frac{2}{9}e^{-5/3}\sum_n \left[7+2n\right](-x)^n
=\frac{2}{9}e^{-5/3} \left[7-\frac{2x}{1+x}\right]\frac{1}{1+x},
\eeq
where $x=\beta b$ and the factor $e^{-5/3}$ transforms the result from the MOM
into the $\overline{\rm MS}$ scheme. We observe that
the leading UV renormalon is a double pole. Since in the large-$\beta$
limit $\beta^\prime$ can be neglected, the corresponding
expression for $R_\tau$ in the approximation of eq.~(\ref{rapp})
is appropriate, and it turns the double pole into a simple pole.

We now study the numerical effect of including the whole renormalon series
with respect to a truncated result up to the order $b^2$.
We first consider the impact on $d(a)$, i.e. before the integration
on the circle. In order to get the correction to $d(a)$ from the higher-order
terms in the UV renormalon we must subtract from $B(b)$ its expansion up
to ${\cal O}(b^2)$ and perform the inverse Borel transform,
eq.~(\ref{antiborel}):
\beq
\Delta d(a)=\frac{1}{\beta}\int_0^\infty dx \,e^{-x/\beta a}
\left[B(x)-\frac{2}{9}e^{-5/3}(7-9x+11x^2)\right].
\eeq
For $\beta=27/12=2.25$ and $a=0.12$ (or $\beta a$ = 0.27) one finds
$\Delta d \sim - 3.7\times 10^{-3}$ which corresponds to a $\sim$2\%
increase in the value of $\as(m^2_\tau)$ at
fixed $d(0.12)=0.155$ ($\delta\as(m^2_\tau)\sim 0.007$).

We now repeat the same exercise for the function $r$, given in
eq.~(\ref{rapp}), obtained after integration over the circle. We
compute the variation
\beq
\Delta r(a)=\frac{1}{\beta}\int_0^\infty dx \,e^{-x/\beta a}
\left[B(x)-\frac{2}{9}e^{-5/3}(7-9x+11x^2)\right]F(x).
\eeq

Numerically we find $\Delta r(0.12) \sim -6.0 \times 10^{-3}$ which,
at fixed $r(0.12) =0.220$, again corresponds to
$\delta\as(m^2_\tau)\sim 0.007$. Thus the extra factor
$F(\beta b)$ has practically no influence on the effect on $\as(m^2_\tau)$
of the nearest UV singularity.

In conclusion the overall effect of the UV renormalon singularity in
this model is small and not much changed by the integration over
the circle.

\section{Search for More Convergent Approximants}
Assuming that indeed there is no IR renormalon at $b=+1/\beta$ one can in
principle try to obtain by analytic continuation a definition of the
Borel transform, valid on the positive real $b$ axis up to $b=+2/\beta$,
outside the radius of convergence of its expansion. We now discuss how
the analytic continuation could be implemented in practice.

Starting from eq.~(\ref{boreltr}) we can make a change of variable
[\ref{Lautrup},\ref{Mueller1}] $z=z(b)$ with inverse $b=b(z)$, $z(0)=0$
and $z(\infty)=1$ (so that the interval
from 0 to $\infty$ in $b$ is mapped into the 0 to 1 range in $z$),
such that the IR singularities are mapped onto the interval between
$z_0=z(2/\beta)$ and 1 and the UV singularities are pushed away at
$|z|\geq z_0$.
Changing variable one obtains
\beq
d(a)=\int_0^\infty db \, e^{-b/a} B(b)
\int_0^1 dz \abs{\frac{db}{dz}} e^{-b(z)/a} B(b(z)).
\eeq
Using the expansion
\beq
b(z) = c_1 z + c_2 z^2+\ldots
\eeq
the series
\beq
B(b) = D_1 +D_2 b + D_3 \frac{b^2}{2}+\ldots
\eeq
goes into
\beq
B(b(z)) = D_1 + D_2c_1z + (D_2c_2 + D_3\frac{c_1^2}{2})z^2+\ldots
\eeq
which is convergent up
to $z=z_0$, while the original $b$ expansion was convergent only up
to $b=1/\beta$, corresponding to $z(1/\beta)<z_0$.
The improved approximation for $d(a)$ is therefore given by
\beqn
d(a) &\simeq&
\int_z^{z_0} dz
\abs{\frac{db}{dz}} e^{-b(z)/a}
\left[D_1 + D_2c_1z + (D_2c_2 + D_3\frac{c_1^2}{2})z^2+\ldots\right]
\nonumber\\
&=&
\int_0^{2/\beta} db \, e^{-b/a}
\left[D_1 + D_2c_1z(b) + (D_2c_2 + D_3\frac{c_1^2}{2})z(b)^2+\ldots\right],
\eeqn
where the full expression of $z$ as function of $b$ is inserted in the
integral. In this way, an infinite sequence of terms is added to the $b$
expansion. For $a$ small, the upper limit of integration can be replaced
with infinity without significant effect.

One possible example is given by [17]:
\beq
\label{Macc}
z(b)=\frac{\sqrt{1+\beta b}-1}{\sqrt{1+\beta b}+1}
\;\;\to\;\;
b(z)=\frac{4z}{\beta(1-z)^2}.
\eeq
In this case the first UV singularity is at $z = -1$, and all higher UV
renormalons are on the unit circle $|z| = 1$. IR renormalons are between
$z_0=(\sqrt{3}-1)/(\sqrt{3}+1)$ and $z=1$. In this example,
$c_1=4/\beta$, $c_2=8/\beta$. Other examples are
\beq
\label{Kacc}
z(b)=\frac{\beta b}{k+\beta b}
\;\;\to\;\;
b(z)=\frac{kz}{\beta(1-z)},
\eeq
with $k= 1,2$ or 3. Also in these cases the first IR renormalon at
$b =2/\beta$ becomes the closest singularity to $z = 0$, while the UV
are pushed further away. Here we have $c_1 =  c_2 = k/\beta$.

Before discussing numerical applications, we observe that the present
method relies simply on the position of the IR and UV renormalon
singularities in the Borel plane and not on the nature and the
strength of the singularities. We have seen that the integration over
the circle in eq.~(\ref{rapp}) does not change the position of the
singularities
in the $b$ plane, but simply affects their strength. Thus, we can as
well consider the effect of the accelerators on the
expansions for $R_\tau^{({\rm BNP})}$ given in
eq.~(\ref{rtauBNPscale}) or on the LP expression of
eq.~(\ref{rtaupichmu}).
For example, the improved version of eq.~(\ref{rapp}) simply becomes
\beq
r\simeq\int_0^{\infty} db \, e^{-b/a}
\left[D_1 + D_2c_1z(b) + (D_2c_2 + D_3\frac{c_1^2}{2})z(b)^2+\ldots\right]
F(\beta b).
\eeq

\begin{figure}[tbhp]
  \begin{center}
    \mbox{
      \epsfig{file=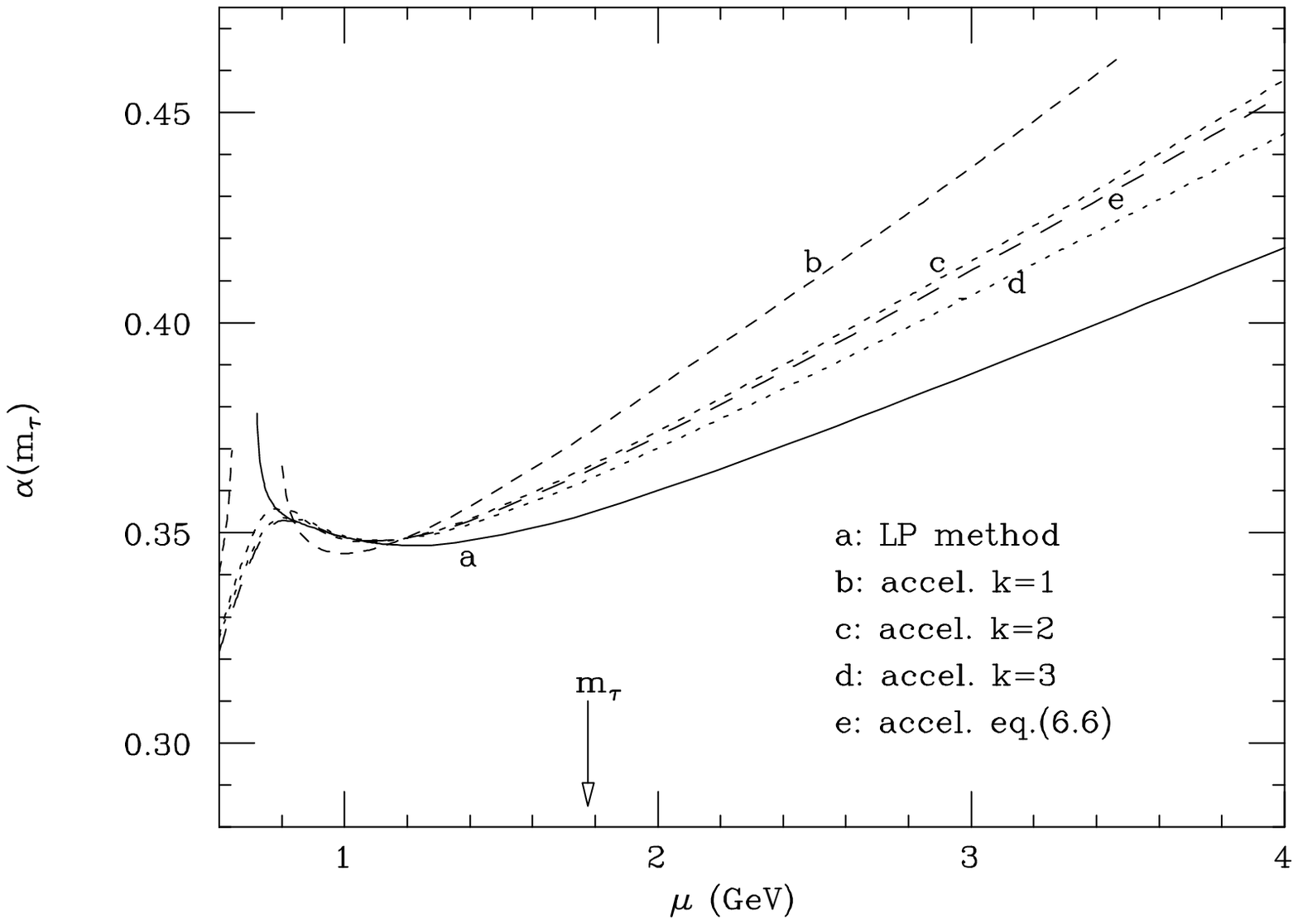,width=0.80\textwidth}
      }
\settmpfig{Kacc}
\settttfig{Macc}
  \ccaption{}{\label{rtau_exact_improv}
  Effect of the accelerators on the determination of $\as(m^2_\tau)$
  with the LP method, for $R_\tau=3.6$. The curves b, c and d refer to
  the change of variable of eq.~(\ref{Kacc}), while e refers to
  eq.~(\ref{Macc}).
   }
  \end{center}
\end{figure}

\begin{figure}[tbhp]
  \begin{center}
    \mbox{
      \epsfig{file=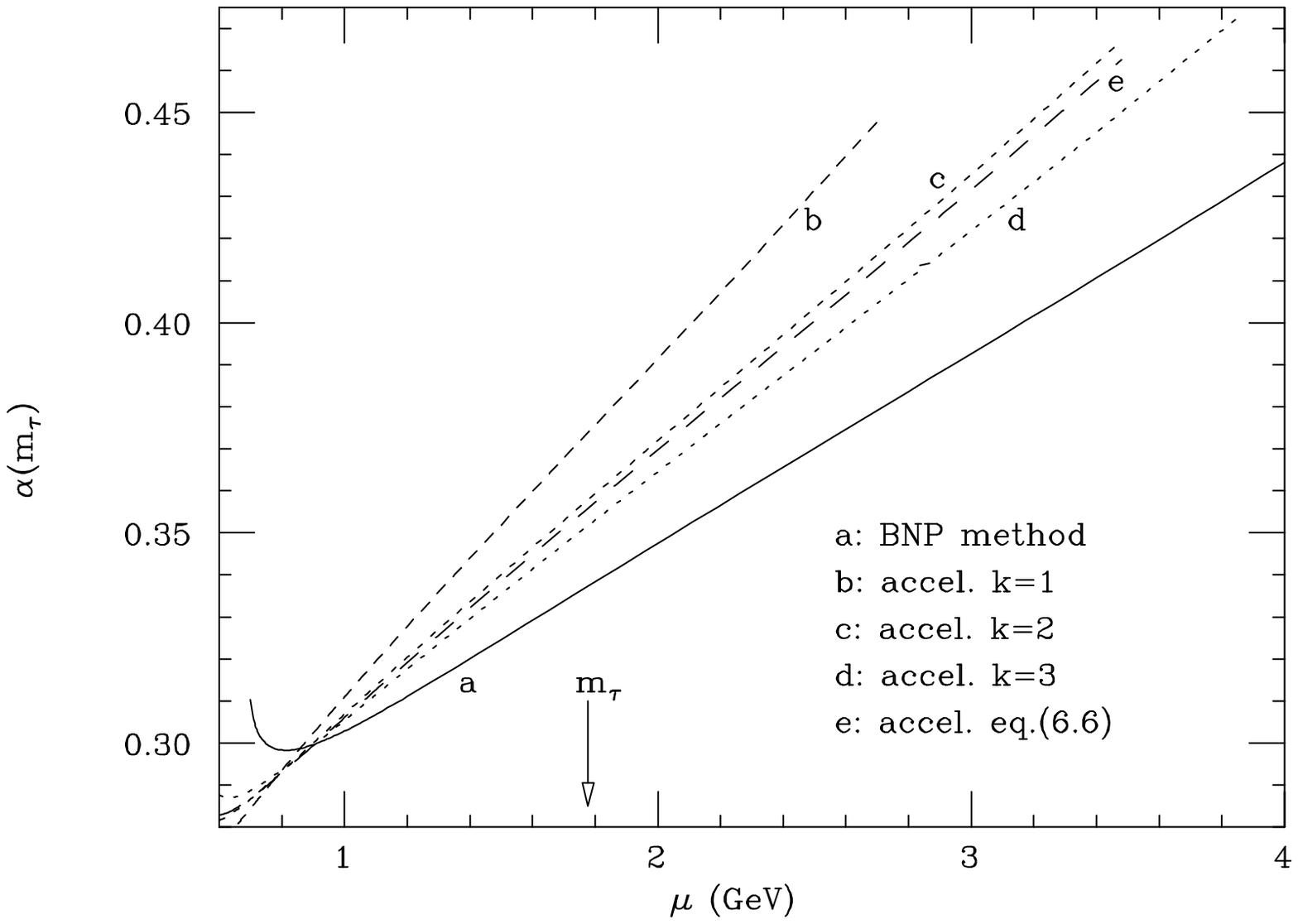,width=0.80\textwidth}
      }
\settmpfig{rtau_exact_improv}
  \ccaption{}{\label{rtau_braaten_improv}
  As in Fig.~\box\tmpfig, for the BNP method.
   }
  \end{center}
\end{figure}

We now consider the following numerical exercise. We assume
that experiments have measured $R_\tau=3.6$.
We then compute $a(m^2_\tau)$ with the LP formula
as a function of the scale $\mu$, and we perform the same calculation
applying our acceleration procedures to the LP method [\ref{DiberderPich}].
The results are shown in fig.~\ref{rtau_exact_improv}.
We see that relatively large differences in the
fitted value of $\as(m^2_\tau)$ are obtained, especially at large $\mu$ for
different accelerators and in comparison to the non-accelerated
formulae. We do not see a priori compelling reasons to prefer one or
the other procedure. The fact that a priori equivalent methods lead
to results with a sizeable spread must be considered as an
indication of a real ambiguity. Even if we only consider the method
of ref.~[\ref{DiberderPich}] for the integration over the circle, it is
impossible to go below an uncertainty of the order
$\delta\as(m^2_\tau)\sim \pm 0.050$ for $\mu$ in the range from 1 to 3~GeV.
The ambiguity becomes even larger if we extend the comparison to the formulae
with truncation in $\beta^2a^2\pi^2$ a--la BNP
(fig.~\ref{rtau_braaten_improv}).

\section{Study of More Convergent Approximants in a Model}
The method for accelerating the convergence discussed in the previous
section only relies on the position of the singularities of the Borel
transform and not on their nature and strength. It looks rather
surprising that one can compensate for the effect of renormalons
without actually knowing their form in detail. In this section we
study a simple mathematical model to clarify under which conditions the
method can be successful, in the sense that it provides a better
approximation to the true result.

We consider as a model the case where the Borel transform is exactly
specified by
\beqn
B_{true}(b)&=&1+D_2 b+D_3\frac{b^2}{2}
+\rho\sum_{n=3}^{\infty}
\left(
\begin{array}{c}
n+\gamma-1\\
   n
\end{array}
\right) \
(-\beta b)^n
\nonumber \\
&=&
1+D_2 b+D_3\frac{b^2}{2}+\rho\left[\frac{1}{(1+\beta b)^\gamma}
-1+\gamma\beta b-\frac{\gamma(\gamma+1)}{2}(\beta b)^2\right].
\label{btrue1}
\eeqn
The added sum stands for the higher-order contribution that could arise from
a leading UV renormalon at $b=-1/\beta$ with a degree of singularity
specified by $\gamma$ and a fixed overall strength given by $\rho$. We will
take $\rho=1$ in the following discussion. It is convenient to re-express
eq.~(\ref{btrue1}) in terms of $x=\beta b$:
\beq
B_{true}(x)=1+\overline{D}_2 x+\overline{D}_3\frac{x^2}{2}+
\left[\frac{1}{(1+x)^\gamma}
-1+\gamma x-\frac{\gamma(\gamma+1)}{2}x^2\right].
\eeq
where $\overline{D}_2=D_2/\beta$ and $\overline{D}_3=D_3/\beta^2$. Similarly
we can introduce $B_{pert}(x)$ and $B_{accel}(x)$, the perturbative Borel
functions without and with acceleration, respectively:
\beq
B_{pert}(x)=1+\overline{D}_2 x+\overline{D}_3\frac{x^2}{2}
\eeq
\beq
B_{accel}(x)=1+\overline{D}_2 \overline{c}_1 z(x)+
(\overline{D}_2 \overline{c}_2+\overline{D}_3\frac{\overline{c}_1^2}{2})z(x)^2,
\eeq
where $\overline{c}_{1,2}=\beta c_{1,2}$. In all cases the corresponding $d$
function, $d_{true}$, $d_{pert}$ and $d_{accel}$ is given by
\beq
\beta d(a)=\int_0^\infty dx \, e^{-x/\beta a} B(x).
\eeq

\begin{figure}[tbhp]
  \begin{center}
    \mbox{
      \epsfig{file=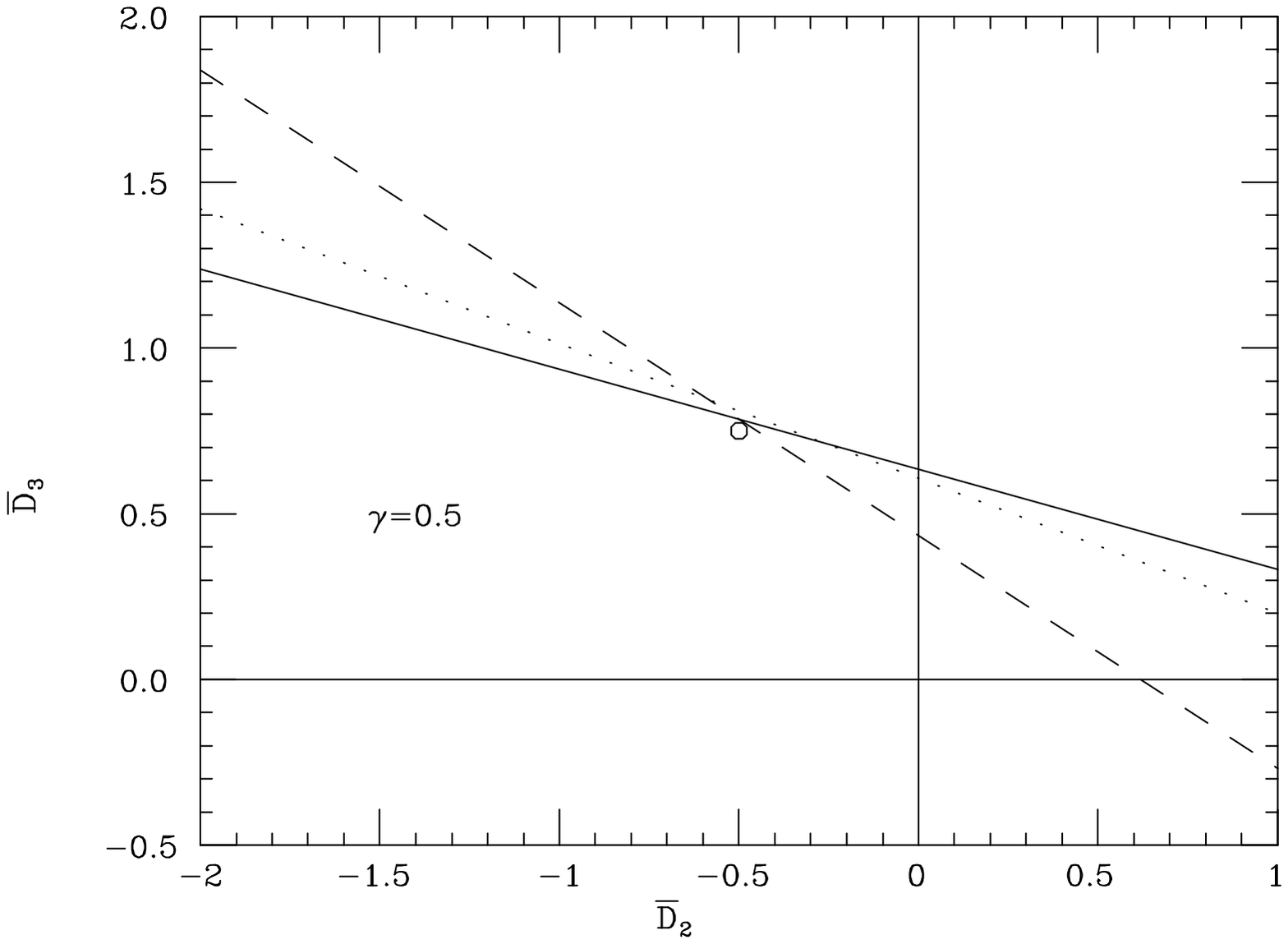,width=0.80\textwidth}
      }
  \ccaption{}{\label{fig5}
  Lines corresponding to $H=0$ for the method ``0'' (solid),
  ``1'' (dashed) and ``2'' (dotted) for $\gamma=0.5$.
  The circle corresponds to expansion of $1/(1+x)^\gamma$.
   }
  \end{center}
\end{figure}

\begin{figure}[tbhp]
  \begin{center}
    \mbox{
      \epsfig{file=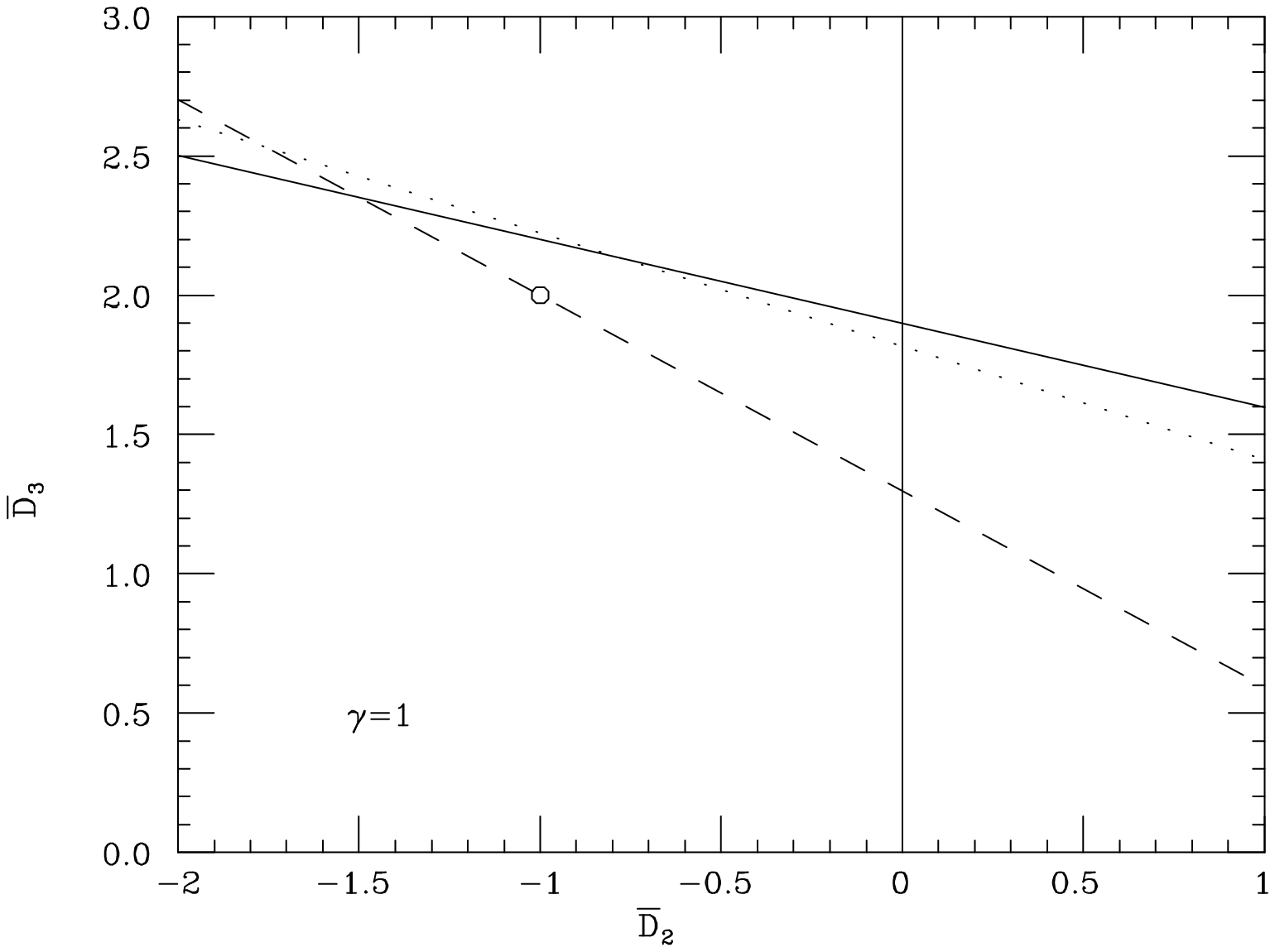,width=0.80\textwidth}
      }
\settmpfig{fig5}
  \ccaption{}{\label{fig6}
  As in Fig.~\box\tmpfig, for $\gamma=1$.
   }
  \end{center}
\end{figure}

\begin{figure}[tbhp]
  \begin{center}
    \mbox{
      \epsfig{file=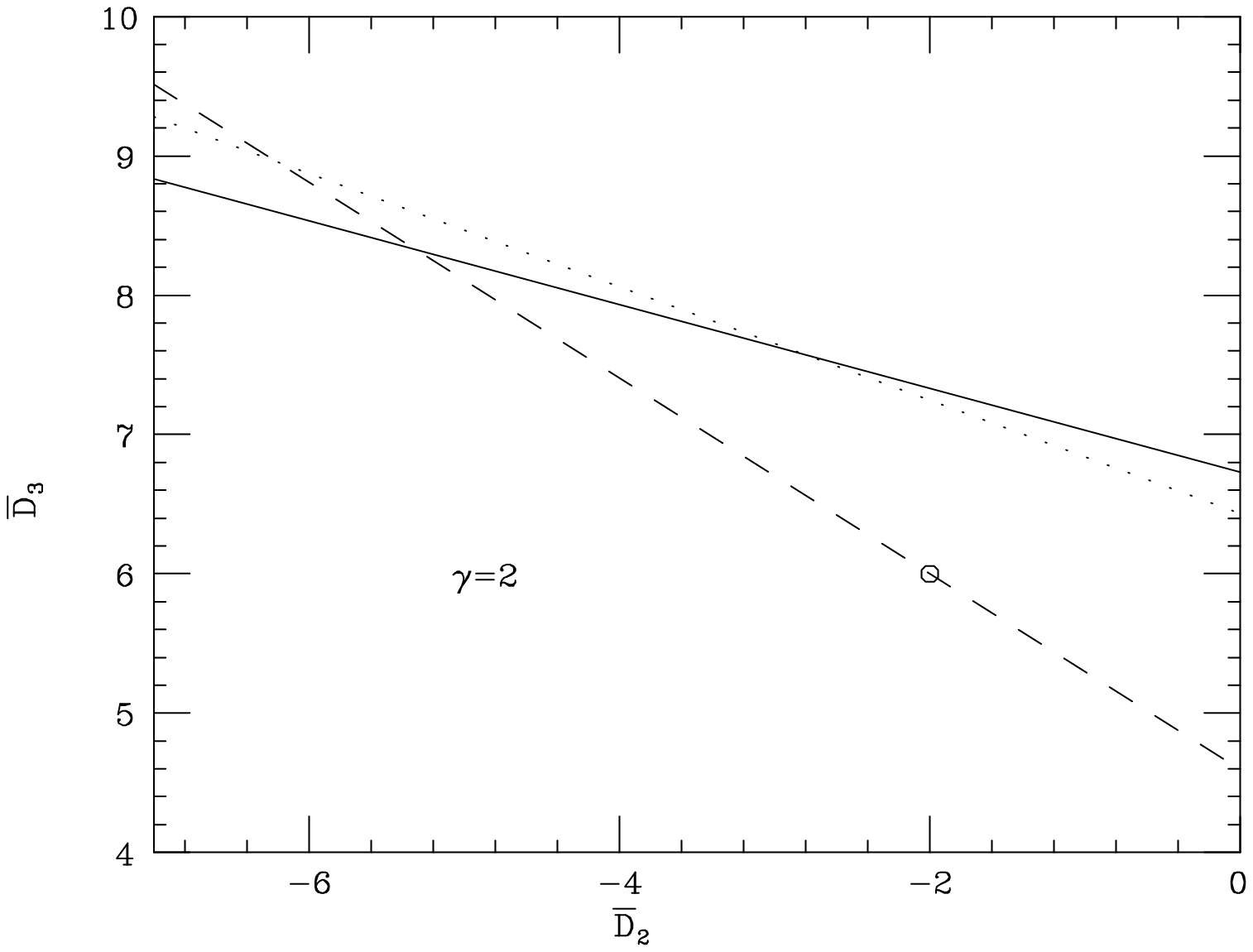,width=0.80\textwidth}
      }
\settmpfig{fig5}
  \ccaption{}{\label{fig7}
  As in Fig.~\box\tmpfig, for $\gamma=2$.
   }
  \end{center}
\end{figure}

\begin{figure}[tbhp]
  \begin{center}
    \mbox{
      \epsfig{file=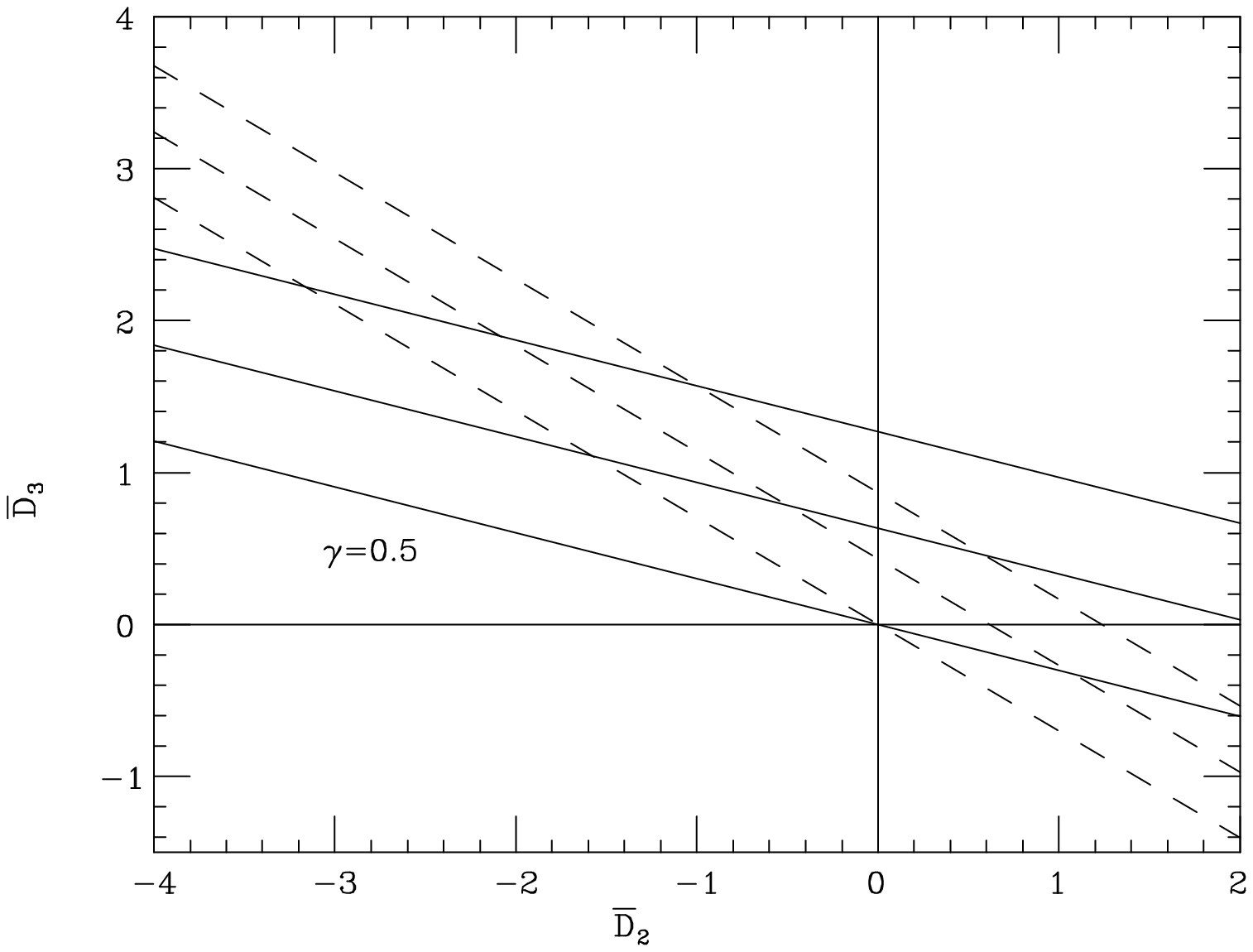,width=0.80\textwidth}
      }
  \ccaption{}{\label{fig8}
  Bands corresponding to $|H|<1$ for the method ``0'' (solid),
  ``1'' (dashed) for $\gamma=0.5$.
   }
  \end{center}
\end{figure}

\begin{figure}[tbhp]
  \begin{center}
    \mbox{
      \epsfig{file=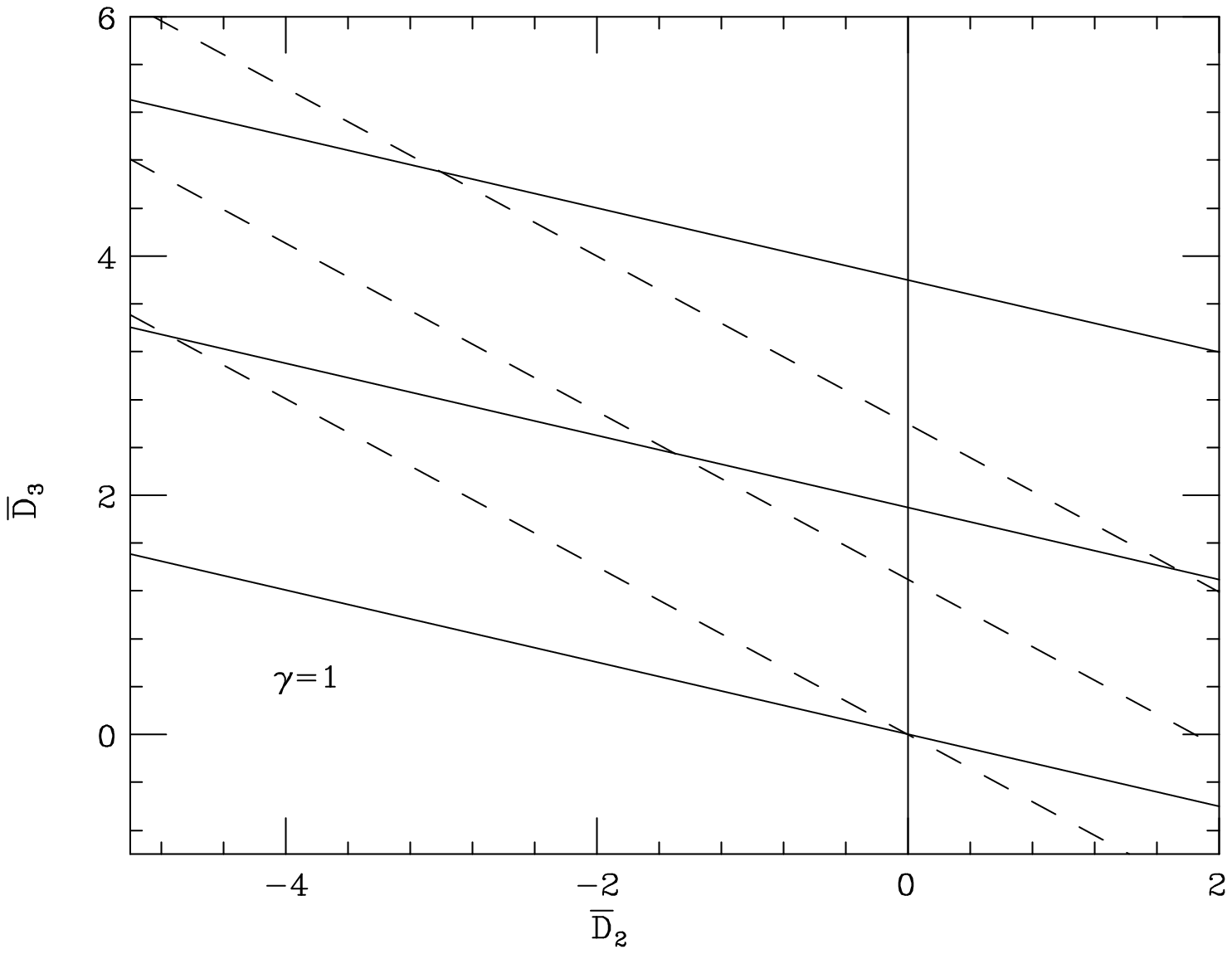,width=0.80\textwidth}
      }
\settmpfig{fig8}
  \ccaption{}{\label{fig9}
  As in Fig.~\box\tmpfig, for $\gamma=1$.
   }
  \end{center}
\end{figure}

We consider the ratio
\beq
H=\frac{d_{true}-d_{accel}}{d_{true}-d_{pert}}
=1-\overline{D}_2 I_2(\beta a)-\overline{D}_3 I_3(\beta a),
\eeq
where the quantities $I_{2,3}$ are given by
\beqn
I_2(\beta a) &=& \frac{1}{I_0(\beta a)}
\int_0^\infty dx\, e^{-x/\beta a}\left(\overline{c}_1 z(x)+
\overline{c}_2 z(x)^2-x\right)
\\
I_3(\beta a) &=& \frac{1}{2}\frac{1}{I_0(\beta a)}
\int_0^\infty dx\, e^{-x/\beta a}\left(\overline{c}_1^2 z(x)^2-x^2\right)
\eeqn
and
\beq
I_0(\beta a)=\int_0^\infty dx\,e^{-x/\beta a}
\left[\frac{1}{(1+x)^\gamma}
-1+\gamma x-\frac{\gamma(\gamma+1)}{2}x^2\right].
\eeq
Clearly, $|H|<1$ is the condition for the acceleration method
to be successful. In particular for $H=0$ $d_{true}$ and $d_{accel}$
coincide. For each value of $\beta a$ and $\gamma$,
in a given model specified by $z(x)$ and the corresponding coefficients
$\overline{c}_{1,2}$, the condition $H=0$ is satisfied
on a straight line in the
plane $\overline{D}_2,\overline{D}_3$,
while the inequality $|H|<1$ is satisfied in
a band defined by two straight lines parallel to the $H=0$ line.
In figs.\ref{fig5}-\ref{fig7} we plot the lines $H=0$ for $\beta a=0.27$
for fixed $\gamma$ and $z(x)$ given by eq.~(\ref{Macc}) (case labelled by 0),
or by eq.~(\ref{Kacc}) with $k=1$ or 2 (cases 1 and 2).
The values of $\gamma$ in figs.~\ref{fig5}-\ref{fig7}
are $\gamma=0.5,1,2$. We see that for each choice of $z(x)$ the lines have
different negative slopes. The lines tend to cross each other in a
region of the plane not far from the point $\overline{D}_2=-\gamma$,
$\overline{D}_3=\gamma(\gamma+1)$, i.e. the values that correspond to the first
few terms of the expansion of the
asymptotic function  $1/(1+x)^\gamma$. The region where the lines cross is
more sharply defined if $\gamma$ is small, i.e. if the asymptotic function is
not too singular. In figs.~\ref{fig8}--\ref{fig9} we show the bands
$|H| < 1$ for  $\gamma = 0.5,1$ in cases 0 and 1.

The conclusion is that the method for accelerating the
convergence works well only if the coefficients
$\overline{D}_2,\overline{D}_3$ resemble those of the asymptotic series,
in other words if the known terms in the expansion are
sufficiently representative of the asymptotic series. In particular we
see that it is very unlikely to get an improvement if the coefficients
$\overline{D}_2$ and $\overline{D}_3$ are of the same sign, as is unfortunately
the case for the series of interest for us (see eq.6).

\begin{figure}[tbhp]
  \begin{center}
    \mbox{
      \epsfig{file=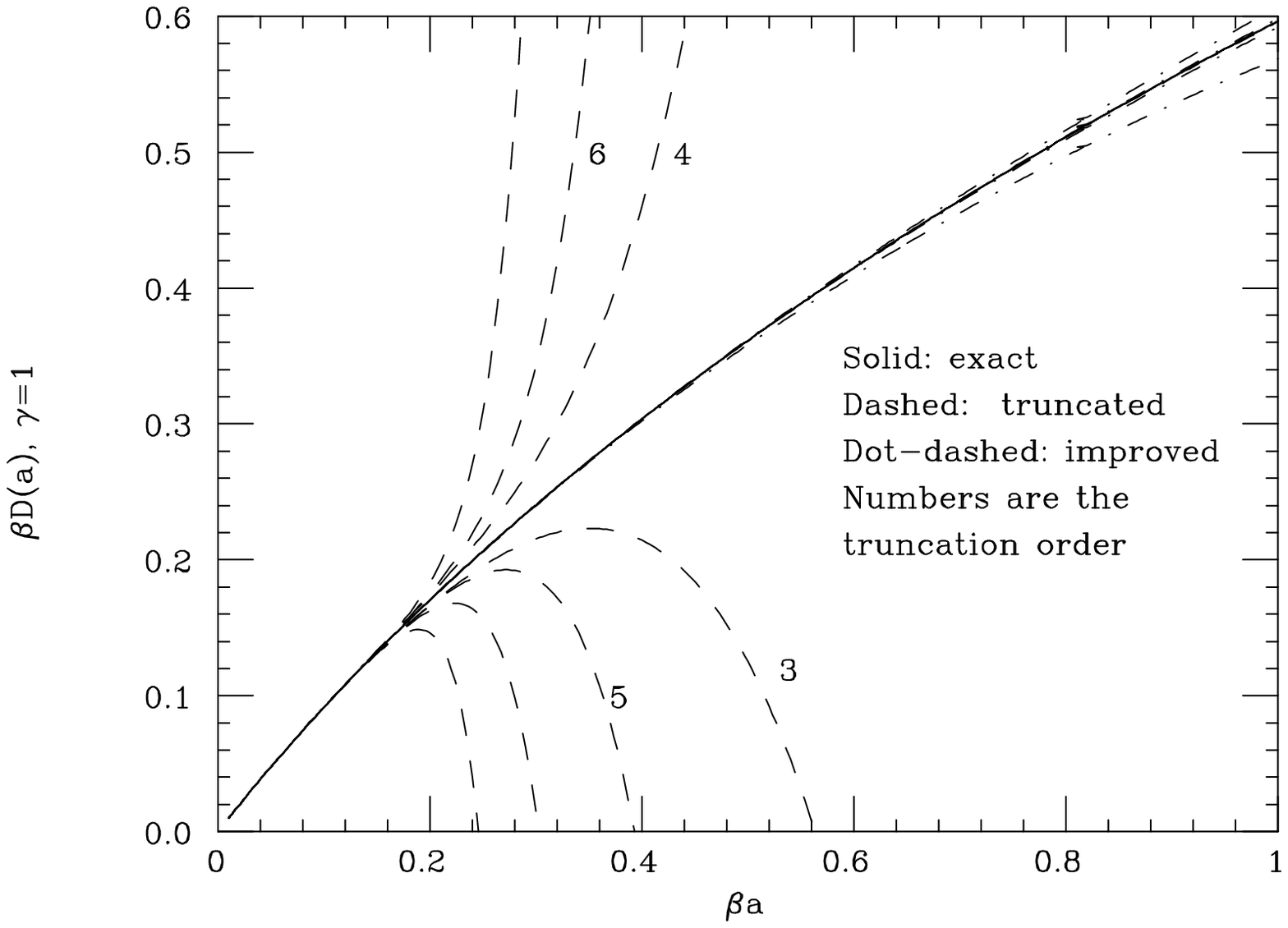,width=0.80\textwidth}
      }
  \ccaption{}{\label{improved_example_1}
Effect of the resummation technique described in the text, for
a function with Borel transform $B(b)=1/(1+\beta b)^\gamma,\gamma=1$}
  \end{center}
\end{figure}

\begin{figure}[tbhp]
  \begin{center}
    \mbox{
      \epsfig{file=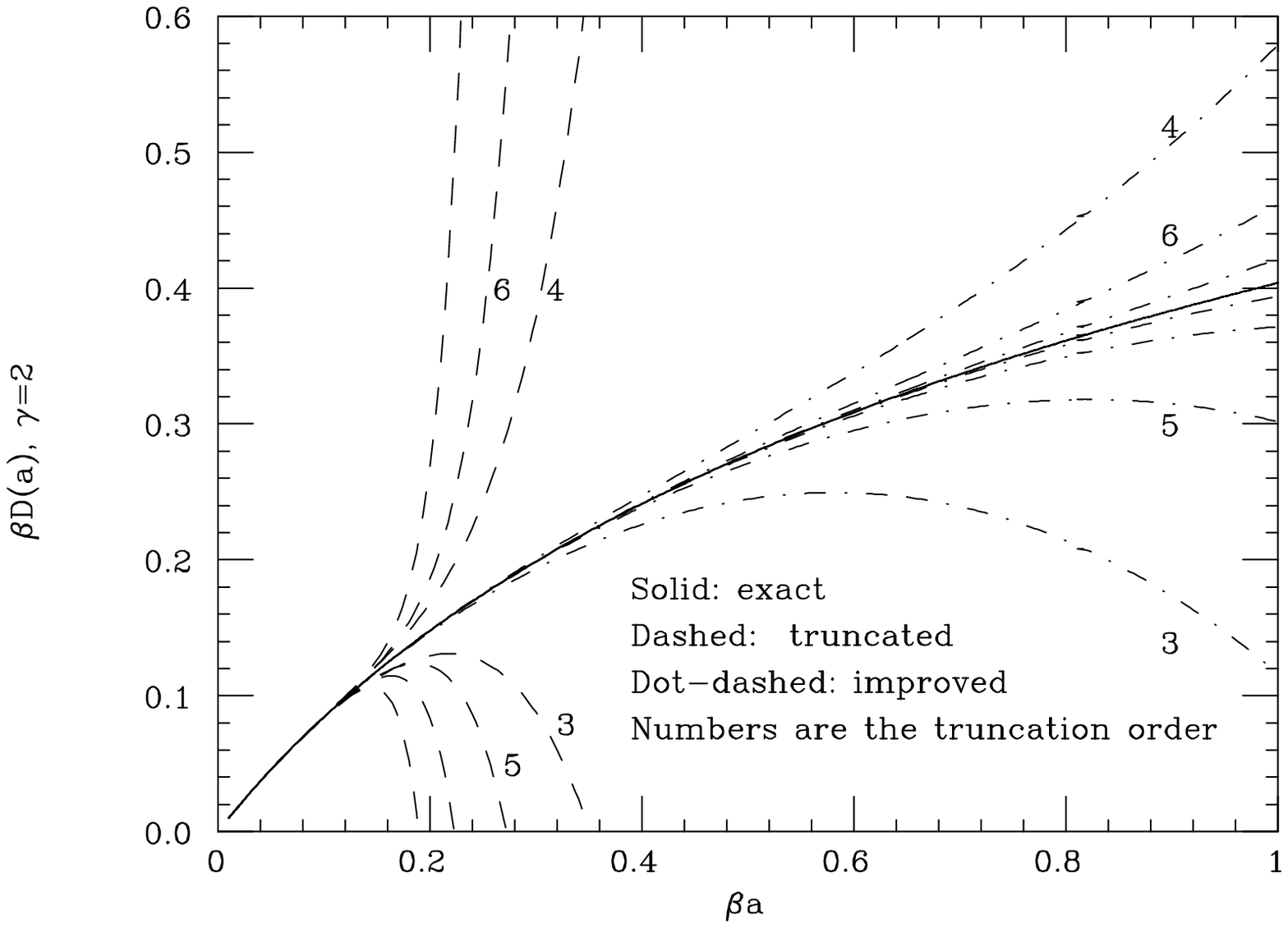,width=0.80\textwidth}
      }
  \ccaption{}{\label{improved_example_2}
Effect of the resummation technique described in the text, for
a function with Borel transform $B(b)=1/(1+\beta b)^\gamma,\gamma=2$}
  \end{center}
\end{figure}

{}From a different point of view we now consider the simple
function $B(\beta b)$ given by
\beq
B(\beta b)=\frac{1}{(1+\beta b)^\gamma}
\eeq
and we plot the relation of the exact result
\beq
\beta d(a)=\int_0^\infty dx \, e^{-x/\beta a} B(x)
\eeq
with its accelerated or non-accelerated series approximants, as a
function of $\beta a$ and of the order of the expansion. The results obtained
for the accelerating function $z(b)$ given in eq.~(\ref{Macc})
(the case labeled by 0) and $\gamma = 1,2$ are shown in
figs.~\ref{improved_example_1},\ref{improved_example_2}.
We see that when, as in this
case, the coefficients of the expansion coincide with their asymptotic
form the accelerated formulae provide a much better approximation
to the true result, more so if the singularity is weaker (i.e. $\gamma$ is
smaller). The non-accelerated formulae are only good for small enough
$\beta a$. The physically interesting case of $\beta a\sim 0.27$ appears
to be at the limit of the range where the non-accelerated formulae
are acceptable.

\section{Conclusion}
The determination of $\as(m^2_\tau)$ or $\as(m^2_{\sss Z})$ from $\tau$ decay
is
nominally very precise and the experimental errors are extremely small
at LEP. Certainly the dominant ambiguity is at present the theoretical
error. The nominal precision is large because in the massless limit no
explicit $1/m^2_\tau$ corrective terms are present in the operator expansion.
But it has become clear by now that one cannot sensibly talk of
power-suppressed
corrections if the ambiguities in the leading term are not under
control [\ref{Mueller2}].
We think that there is no real theorem that prevents non-perturbative
corrections in the coefficient function of the leading
term in the operator expansion at the level of $\Lambda^2/m_\tau^2$.
Equivalently, there could be an IR renormalon singularity at
$b=+1/\beta$, which would create an irreducible (being located on the
integration path) ambiguity of order $\Lambda^2/m_\tau^2$. In all-order
perturbative evaluations of the singularity pattern in the Borel plane the IR
singularity at $b=+1/\beta$ is probably
absent, a result consistent with the idea that all irreducible
ambiguities can be reabsorbed in condensates. Even if the IR
renormalon singularity at $b=+1/\beta$ is indeed absent, the radius of
convergence of the expansion is limited by the leading UV renormalon
singularity at $b=-1/\beta$. If this disease is not cured or cannot be cured
the resulting ambiguity is still of order
$\Lambda^2/m_\tau^2$. In principle the problem could be solved if the exact
nature and strength of the
singularity was known, by simply taking its effect into account in the
evaluation of $\as(m^2_\tau)$, along the way indicated in section 5 in the case
of the estimate of the singularity in the unrealistic limit of large
$\beta$. But the exact determination of the singularity appears to be beyond
the scope of presently known methods. In the actual case where the UV
renormalon singularity at $b=-1/\beta$ is not specified, one can still try,
in principle, to bypass the problem by a transformation of variables
that pushes the leading UV singularity to a larger distance from the
origin than the first IR renormalon singularity at $b=+2/\beta$.
Expanding in the new variable is equivalent to add a specified infinite sequel
of terms to the original expansion. The convergence of the series
should be improved by these accelerators of convergence and the
ambiguity decreased. We have studied the quantitative effect of
implementing a number of such accelerators with different choices of
the renormalisation scale $\mu$. An indication of the size of the
ambiguities on $\as(m^2_\tau)$ is obtained from the spread of the results for
different starting formulae (e.g. with $a(-s)$ taken in the integration
over the circle in its renormalisation group improved form or in a
fixed order truncated expansion), different accelerator functions and
different choices of the renormalisation scale. The difference between
the resummed or truncated expression for $a(-s)$ on the circle are
expecially large at small values of $\mu$, while the variations induced by
the different accelerators are expecially pronounced at large values
of $\mu$. The relative stability of the unaccelareted result of
ref.~[\ref{DiberderPich}] versus changes of $\mu$ appears as largely
accidental in that the
accelerated formulae based on it are much less stable at large $\mu$. It
was argued in ref.~[\ref{Beneke}] that if one expands in $\as(\mu^2)$
instead of expanding in
$\as(m^2_\tau)$ the scale dependence of the UV renormalon correction becomes of
order $(\Lambda^2/m_\tau^2)(m_\tau^2/\mu^2)^2$. Can then one be safe
if $\mu$ is chosen sufficiently large? Clearly in the true
result the sum of the perturbative terms plus the remainder
must be scale independent. When $\mu$ is changed, the number of terms to
be added before the series becomes asymptotic changes and must
compensate for the difference. The increased sensitivity of the
accelerated formulae at large $\mu$ is not encouraging for invoking that
large $\mu$  is safer. All together, from fig.4 we find it difficult to
imagine that the theoretical error on the strong coupling can be taken
smaller then, say, $\delta\as(m^2_\tau)\sim 0.050$ (which approximately
corresponds to $\delta\as(m^2_{\sss Z}) \sim 0.005$).

The accelerator method is based on the mere knowledge of the position of the
singularity and not on its precise form. Clearly such a method can
only work if the known terms of the expansion carry enough information
on the asymptotic form of the series. We have quantitatively confirmed
this statement by studying the problem on a simple mathematical model
where the true result is known. The performance of different
accelerators is studied as a function of the coefficients of the first
few terms. These results indicate that there is little hope of
improving the ambiguity from the leading UV renormalon because the
first few coefficients of the actual expansion show no evidence for
the asymptotic behaviour, in particular no sign alternance. This last
argument (as well as the one on the $\mu$ dependence of the UV renormalon)
can be interpreted in different ways. If one is a great
optimist, he can argue that the series does not resemble at all to the
renormalon asymptotics, hence the normalisation of the renormalon term
is very small (as is the case for the explicit form of the singularity
obtained in the large $N_f$ limit). Or, if one is more cautious, as one
should be in estimating errors, he can say that since the known terms
do not show sign of asymptotia, they are dominated by subasymptotic
effects and cannot be used to estimate the remainder. In this spirit
we do not propose the accelerators as a better way to determine the
true result but simply as a criterium to evaluate the theoretical
error. In fact, while all accelerators tend to increase the resulting
value of $\as$, the amount of the upward shift is sizeably different for
different accelerators.

In order to bypass all possible objections one should be able
to fit at the same time $\as(m^2_\tau)$ and $C_2$, the coefficient of
$1/m^2_\tau$  corrections to $R_\tau$. Note that in the ALEPH moment
analysis [\ref{Duflot}] $C_2$ is fixed to zero while the coefficients
of some higher-dimension operators are fitted. This is not very relevant to
the main issue. If $C_2$ is not fixed it is found that the sensitivity to
$\as(m^2_\tau)$ is much reduced. In an interesting paper
Narison [\ref{Narison2}]
attempted to put an upper bound on $C_2$  from the data on
$e^+e^- \to$~hadrons. This is an important issue that would deserve further
study. Our interpretation of the analysis of ref.~[\ref{Narison2}] is that
values of $C_2$ of order (500 MeV)$^2$ are not at all excluded. Narison
[\ref{Narison2}] derives a more stringent limit $|C_2|<(374$ MeV$)^2$
but we feel he relies too much on the so called
optimisation procedure. Indeed, something that should be a constant in a dummy
variable turns out to be a steep parabola. The value at the tip is
taken, with a small error, as the best estimate because of the
vanishing of the derivative at that point, instead of considering the
span of the results in a priori reasonable range for the irrelevant
parameter. In a recent paper [\ref{Dominguez}] an estimate of $C_2$
from Argus data on hadronic $\tau$ decay was obtained and the results are
compatible with $|C_2| < (500$~MeV$)^2$.

We ignored here other possible sources of error beyond those
arising from higher orders in perturbation theory. These include
errors from the freezing mechanism for $\as$, errors from the translation
of $\as(m^2_\tau)$ in terms of $\as(m^2_{\sss Z})$, from the region of the
circle
integration near the positive real axis and so on. These errors are
presumably smaller [\ref{GA2}] than our current estimate of the error from
higher order terms in the perturbative espansion. Taking all the other
uncertainties into account we end up with a total theoretical error
around $\delta\as(m^2_{\sss Z}) \sim 0.006$.
As a result, in spite of the fact that our estimate of the error is larger
than usually quoted, the determination of $\as(m^2_{\sss Z})$ from $\tau$
remains one of the best determinations of the strong coupling constant.

\begin{reflist}
\item\label{SchilcherTran}
    K.~Schilcher and M.D.~Tran, \pr{D29}{84}{570}.
\item\label{Braatenetal}
    E.~Braaten, \prl{60}{88}{1606}; \\
    S.~Narison and A.~Pich, \pl{B211}{88}{183}; \\
    E.~Braaten, \pr{D39}{89}{1458}; \\
    E.~Braaten, S.~Narison and A.~Pich, \np{B373}{92}{581}.
\item\label{LuoMarciano}
    M.~Luo and W.J.~Marciano, preprint BNL-47187 (1992).
\item\label{DiberderPich}
    F.~Le Diberder and A.~Pich, \pl{B286}{92}{147}.
\item\label{Duflot}
    L.Duflot, Proceedings of the 3rd Workshop on Tau Lepton Physics,
Montreux, 1994.
\item\label{SVZ}
    M.A.~Shifman, A.I.~Vainshtein and V.I.~Zakharov, \np{B147}{79}{385},
    \np{B147}{79}{448}, \np{B147}{79}{519}.
\item\label{Pumplin}
    J.~Pumplin, \pr{D41}{90}{900}.
\item\label{GA1}
    G.~Altarelli,  in ``QCD-Twenty Years Later'', Aachen 1992,
   eds. P.~Zerwas and H.A.~Kastrup, World Scientific, Singapore (1992).
\item\label{GA2}
    G.Altarelli, Proceedings of the 3rd Workshop on Tau Lepton
    Physics, Montreux, 1994.
\item\label{Truong}
    T.N.~Truong, Preprint Ecole Polytechnique, EP-CPth A266.1093 (1993);\\
    \pr{D47}{93}{3999}.
\item\label{Narison1}
    S.~Narison, Proceedings of the 3rd Workshop on Tau Lepton Physics,
    Montreux, 1994.
\item\label{GKL}
    S.G.~Gorishny, A.L.~Kataev and S.A.~Larin, \pl{B259}{91}{144};\\
    L.R.Surguladze and M.A.Samuel, \prl{66}{91}{560}.
\item\label{tHooft}
    G.'t Hooft, in ``The Whys of Subnuclear Physics'', Erice 1977,
    ed. by A.Zichichi, Plenum, New York.
\item\label{Lautrup}
    B.~Lautrup, \pl{B69}{77}{109};\\
    G.~Parisi, \pl{B76}{78}{65}, \np{B150}{79}{163}.
\item\label{David}
    F.David, \np{B234}{84}{237};\\
    A.H.~Mueller, \np{B250}{85}{327}.
\item\label{West}
    G.B.~West, \prl{67}{91}{1388},{67}{91}{3732} (E);\\
    L.S.~Brown and L.G.~Yaffe, \pr{D45}{92}{398};\\
    L.S.~Brown, L.G.~Yaffe and C.~Zhai, \pr{D46}{92}{4712}.
\item\label{Mueller1}
    A.H.~Mueller, in ``QCD-Twenty Years Later'', Aachen 1992,
    eds. P~Zerwas and H.A.~Kastrup, World Scientific, Singapore.
\item\label{Zakharov}
    V.I.~Zakharov, \np{B385}{92}{452}.
\item\label{Beneke}
    M.~Beneke and V.I.~Zakharov, \prl{69}{92}{2472}; \\
    M.Beneke, \np{B385}{92}{452}.
\item\label{VZ}
    A.I.~Vainshtein and V.I.~Zakharov, University of Minnesota preprint
    TPI-MINN 94/9T (1994).
\item\label{Lovett}
    C.N.~Lovett-Turner and C.J.~Maxwell, University of Durham Preprint
    DPT/94/58.
\item\label{GA3}
    See, for example, G.~Altarelli, {\it Phys. Rep.}{\bf 81}(1982)1.
\item\label{Pennington}
    M.R.~Pennington and G.G.~Ross, \pl{B102}{8}{167}; \\
    G.~Parisi, \pl{B90}{80}{295}; \\
    G.~Curci and M.~Greco, \pl{B92}{80}{175};\\
    A.P.~Contogouris et al, \pr{D25}{82}{1280}; \pr{D28}{83}{1644};
    {\it Int. J. Mod. Phys.}{\bf A5}(1990)1951.
\item\label{maghi}
    P.M.~Stevenson, \pr{D23}{81}{2916};\\
    G.~Grunberg, \pl{B221}{80}{70}; \pr{D29}{84}{2315};\\
    S.~Brodsky, G.P.~Lepage and P.B.~Mackenzie, \pr{D28}{83}{228};\\
    H.J.~Lu, \pr{D45}{92}{1217};\\
    M.~Neubert, preprint CERN-TH.7487/94;\\
    A.L.~Kataev and V.V.~Starshenko, preprint CERN-TH.7198/94;
    CERN-TH.7400/95;\\
    P.A.~Raczka and A.~Szymacha, Warsaw University preprint IFT/13/94,
    hep-ph 9412236.
\item\label{Instantons}
    P. Nason and M. Porrati, \np{B421}{94}{518};\newline
    I.I. Balitsky, M. Beneke and V.M. Braun,
    \pl{B318}{93}{371};\newline
    P. Nason and M. Palassini, CERN-TH.7483/94, hep-ph/9411246.
\item\label{TVZ}
    O.V.~Tarasov, A.A.~Vladimirov and A.Yu.~Zharkov, \pl{B93}{80}{429}.
\item\label{Hardy}
    G.N.~Hardy, ``Divergent Series'', Oxford University Press, 1949.
\item\label{Mueller2}
    A.H.~Mueller, \pl{B308}{93}{355}.
\item\label{Narison2}
    S.~Narison, \pl{B300}{93}{293}.
\item\label{Dominguez}
    C.A.~Dominguez, University of Cape Town Preprint, UCT-TP-221/94.
\end{reflist}
\end{document}